\documentclass{article}
\usepackage{setspace}

\newcommand{\interval}{$\mathcal{T}$}
\font\MyScript=eusm10 at 14pt

\usepackage{graphicx}
\usepackage{epstopdf}
\usepackage{amsthm}

\begin{document}
\large
\singlespacing
\begin{center}
Studies in Astronomical Time Series Analysis.\\
VI. Bayesian Block Representations
\vskip 0.5in

\vskip .1in
Jeffrey D. Scargle\\
Space Science and Astrobiology Division,\\
NASA Ames Research Center

\vskip .1in
Jay P. Norris\\
Physics Department, Boise State University

\vskip .1in
Brad Jackson\\
San Jos\'{e} State
University, Department of
Mathematics and Computer Science,\\
The Center for Applied Mathematics and Computer Science

\vskip .1in
James Chiang\\
Kavli Institute, SLAC
\end{center}

\begin{abstract}
This paper addresses the problem of detecting and
characterizing local variability in time series
and other forms of sequential data. 
The goal is to 
identify and characterize 
statistically significant variations,
at the same time suppressing 
the inevitable corrupting observational errors.
We present a simple nonparametric
modeling technique and an algorithm implementing it---an improved
and generalized version of {\it Bayesian Blocks}
\cite{scargle_v}---that finds 
the optimal segmentation of the data in the 
observation interval.  
The structure of the algorithm allows it to
be used in either a real-time {\it trigger} mode, or a
\emph{retrospective} mode. Maximum likelihood or marginal posterior
functions to measure model fitness 
are presented for events, binned
counts, and measurements at arbitrary times 
with known error distributions. 
Problems addressed
include those connected with data gaps,
variable exposure,
extension to piecewise linear 
and piecewise exponential representations,
multi-variate time series data,
analysis of variance,
data on the circle,
other data modes,
and dispersed data.
Simulations provide evidence that the
detection efficiency for weak signals 
is close to a theoretical asymptotic limit
derived by \cite{adh}.
In the spirit of Reproducible Research
\cite{donoho_rr}
all of the code and data necessary to 
reproduce all of the figures in this paper 
are included as auxiliary material.
\end{abstract}

Keywords: time series, signal detection, triggers,
transients, Bayesian analysis

\clearpage
\tableofcontents

\vskip 0.5in
\begin{quotation}
{\bf ``The line is similar to a length of time,
and as the points are the beginning and end
of the line, so the instants are the
endpoints of any given extension of time.''}
\emph{Leonardo da Vinci, Codex Arundel, folio 190v.},
c. 1500 A.D. \cite{capra}.
\end{quotation}
\clearpage


\section{The Data Analysis Setting}
\label{introduction}

This paper describes a method for 
nonparametric analysis of time series data
to detect and characterize structure localized 
in time.
\emph{Nonparametric} 
methods seek generic representations,
in contrast to fitting of models to the data. 
By \emph{local} structure 
we mean light-curve features 
occupying sub-ranges of the total observation interval,
in contrast to global signals 
present all or most of the time
(\emph{e.g. }periodicities)
for which Fourier, wavelet,
or other transform methods
are more appropriate.
The goal is to separate statistically significant features 
from the ever-present random observational errors.
Although phrased in the time-domain 
the discussion throughout is applicable to 
measurements sequential in
wavelength, spatial quantities, 
or any other other independent variable.

This setting leads to the following desiderata:
The ideal algorithm would 
impose as few preconditions as possible,
avoiding assumptions
about smoothness or shape of the signal that
place {\it a priori} limitations on scales and resolution. The
algorithm should handle arbitrary sampling (\emph{i.e.}, not be
limited to gapless, evenly spaced data) 
and large dynamic ranges in
amplitude, time scale and signal-to-noise.
For scientific data mining applications and for
objectivity, the method should be largely automatic. 
To the extent possible it should
suppress  
observational errors while preserving 
whatever valid information lies in the data.
It should be applicable to multivariate problems. 
It should incorporate 
variation of the exposure or instrumental efficiency during the
measurement, 
as well auxiliary, extrinsic information, \emph{e.g.}
spectral or color information.
It should be able to
operate both retrospectively (analyze all the data after they are
collected) and in a real-time fashion that triggers on the first
significant variation of the signal from its background level.

The algorithm described here 
achieves considerable success 
in each of these desired features.
In a simple and easy-to-use computational
framework it represents the structure in the
signal in a form handy for further analysis and the estimation of
physically meaningful quantities.  
It includes an automatic penalty
for model complexity, thus 
solving the vexing problems 
associated with model comparison
 in general
and 
\emph{determining the order of the model}
in particular. 
It is exact, not a
\emph{greedy}\footnote{This term refers to 
algorithms that greedily make optimal
improvements at each iteration
but are not guaranteed to converge
to a globally optimal solution.} 
approximation as in \cite{scargle_v}.

Versions of this algorithm have
been used in various applications, such as 
\cite{qin,norris1,norris2,wgs}.

The following sections discuss, in turn,
the basis of segmentation analysis (\S\ref{segmentation}), 
the piecewise constant model 
adopted in this work (\S\ref{model}), 
extensions to 
piecewise linear  and piecewise exponential models
(\S\ref{model_linear_exp}),
the types of data that the algorithm can accept
(\S\S\ref{data_modes} and \ref{mixed}), 
data gaps (\S\ref{gaps}),
exposure variations 
(\S\ref{exposure_variations}),
a parameter from the prior on the number of blocks
(\S\ref{ncp_prior}),
generalities of optimal segmentation of data into blocks 
(\S \ref{opt_segmentation}), 
some error analysis (\S\ref{error_analysis}),
a variety of
block fitness functions (\S \ref{fitness_functions}), 
and sample
applications (\S \ref{examples}). 
Appendices present some MatLab \copyright  \ code,
some miscellaneous results,
and details of other data modes, 
including dispersed
data (\S\ref{dispersed_data}).
Ancillary files are available 
providing scripts and data 
in order to reproduce
all of the figures in this paper.

\subsection{Optimal Segmentation Analysis}
\label{segmentation}

The above considerations point toward 
the most generic possible
nonparametric data model, and have motivated 
the development of data segmentation and 
change-point
methods -- see \emph{e.g.} \cite{fitzgerald,scargle_v}. 
These methods represent the signal structure 
in terms of a 
segmentation of the time interval into blocks,
with each block containing 
consecutive data elements
satisfying some well defined criterion.
The \emph{optimal segmentation} is that
which maximizes some 
quantitative expression of the criterion
-- for example the sum over blocks of 
a goodness-of-fit of a simple model
of the data lying in each block.

These concepts and methods can be applied in
surprisingly general, higher dimensional contexts.
Here, however, we concentrate on 
one-dimensional data ordered 
sequentially with respect to 
time or some other
independent variable.  
In this setting segmentation analysis 
is often called \emph{change-point detection}, 
since it implements models in which a signal's 
statistical properties change discontinuously at
discrete times
but are constant in the segments between these 
change-points (see \S\ref{change-points}).


\subsection{The Piecewise Constant Model}
\label{model}

It is remarkable that all of the desiderata 
outlined in the previous section
can be achieved in large degree by 
optimal fitting of a piecewise-constant model 
to the data. 
The range of the
independent variable ({\it e.g.} time) 
is divided into subintervals
(here called
\emph{blocks})
generally unequal in size,
in which the dependent variable ({\it
e.g.} intensity) is modeled as constant within errors.
Of all possible such ``step-functions'' 
this approach yields 
the best one by maximizing some 
goodness-of-fit measure.

Defining the times ending one block
and starting the next 
as \emph{change-points}, 
the model of the whole observation interval
contains these parameters: 
\begin{description}
\item[(1)] $N_{cp}$: the number of change-points
\item[(2)]  $t^{cp}_{k}$: the change-point starting block $k$
\item[(3)]  $X_{k}$: the signal amplitude in block $k$
\end{description}
\noindent
for $k = 1, 2, \dots N_{cp}+1$.\footnote{ There 
is one more block than 
there are change-points:
The first datum is always considered
a change-point, marking the start of the first block,
and is therefore not a free parameter.
If the last datum is a change-point, it denotes
a block consisting solely of that datum.}
The key idea is that the blocks can be treated 
independently, 
in the sense that a block's fitness depends 
on its data only.
Our simple 
model for each block has effectively two parameters.
The first represents the signal amplitude,
and is treated as a nuisance parameter
to be determined after the change-points 
have been located.
The second parameter is 
the length of the interval spanned by the block.
(The actual start and stop times 
of this interval are needed for
piecing blocks together to form 
the final signal representation, 
but not for the fitness calculation.)

\vskip .2in
{\bf How many blocks?}  
A key issue is how to determine 
the number of blocks,
$N_{blocks} = N_{cp} + 1$.
Nonparametric analysis 
invariably involves controlling 
in one way or another the 
complexity of the estimated representation.
Typically such regulation is 
considered a trade-off of bias and variance,
often implemented by adjusting
a smoothing parameter.

But smoothing is one of the 
very things we are trying to avoid.
The discontinuities at the block edges
are regarded as assets, not liabilities to
be smoothed over. 
So rather than smooth we 
influence the number of blocks
by defining a prior distribution for the
number of blocks. 
Adjusting a parameter controlling the
steepness of this prior establishes relative
probabilities of smaller or larger numbers
of blocks.
In the usual fashion for Bayesian model selection 
in cases with high signal-to-noise 
$N_{blocks}$ is determined by the structure of the signal;
with lower signal-to-noise the prior becomes more and
more important.
In short, 
we are regulating not smoothness but complexity,
much in the way that wavelet denoising \cite{donoho_johnstone} 
operates without smoothing over sharp features
as long as they are supported by the data.
The adopted prior and the determination of its
parameter are discussed in \S \ref{ncp_prior} below.

This segmented representation 
is in the spirit of nonparametric approximation
and not meant to imply that we believe the signal is actually
discontinuous. 
The sometimes crude and blocky appearance of this 
model may be 
awkward 
in visualization contexts, but for
deriving physically meaningful quantities it is not. 
Blocky models are
broadly useful in signal processing \cite{donoho} and have several
motivations. Their simplicity allows exact treatment of various
quantities, such as the likelihood. We can optimize or marginalize
the rate parameters exactly, 
giving simple formulas for the fitness function
(see \S \ref{fitness_functions} and 
Appendix C,\S\ref{appendix_c}). 
And in many applications 
the estimated model itself is less important
than quantities derived from it. For example, while smoothed plots
of pulses within gamma-ray bursts make pretty pictures, one is
really interested in pulse locations, lags, amplitudes, widths, rise
and decay times, {\emph{etc}. All these quantities can be
determined directly from the locations, heights and widths of the
blocks --  accurately 
and free of any smoothness assumptions.

\subsection{Piecewise Linear and Exponential Models}
\label{model_linear_exp}

Some researchers have applied segmentation 
methods with other block representations.
For example 
\emph{piecewise linear} models
have been used in measuring similarity among time 
series and in pattern matching \cite{keogh}
and to represent time series generated
by non-linear processes \cite{tong}. 
While such models may seem better than 
discontinuous step functions, 
their improved flexibility is 
somewhat offset by added complexity of the
model and its interpretation.  
Note further that if continuity is
imposed at the change-points,
a piecewise linear model has
essentially the same number of 
degrees of freedom 
as
does the simpler piecewise constant model. 

We mention two such generalizations,
one modeling the signal as linear
in time across the block:
\begin{equation}\label{model_linear}
x( t ) = \lambda  ( 1 + a ( t -t_{\mbox{\small fid}} )  )
\end{equation}
\noindent
and the second as exponential:
\begin{equation}\label{model_exponential}
x( t ) = \lambda  e^{ a ( t - t_{\mbox{\small fid}} ) } \ \ .
\end{equation}
\noindent
In both cases $\lambda$ 
is the signal strength at the fiducial 
time $t_{\mbox{\small fid}}$ and the coefficient $a$
determines the rate of variation over the block.
Such models may be useful 
in spite of the caveats mentioned above 
and the added complexity of the block fitness functions.
Hence we provide some details 
in Appendix C,
\S\S \ref{model_linear_fitness} and \ref{model_exponential_fitness}.

\subsection{Histograms}
\label{histograms}

For event data 
the piecewise constant representation 
can be interpreted as  
a \emph{histogram} of the measured
values -- one in which the bins
are not fixed ahead of time and
are free to be unequal in size 
as determined by the data.
In this context the time order 
of the measurements is irrelevant.
Once one determines
the parameter in the prior 
on the number of bins, \verb+ncp_prior+,
one has an objective histogram procedure 
in which the number, 
individual sizes, and
locations of the bins 
are determined solely and uniquely by the data.

\subsection{Data Modes}
\label{data_modes}


The algorithms developed here can be used
with a variety of types of data, 
often called \emph{data modes} in 
instrumentation contexts.
An earlier paper \cite{scargle_v}
described several, 
with formulas for the corresponding
fitness functions. 
Here we discuss data modes
in a broader perspective.
It is required that the measurements 
provide sufficient information 
to determine which block they belong
to and then 
to compute the model fitness function
for the block (cf. \S \ref{blocks}).

Almost any physical variable
and any measurement scheme for it,
discrete or quasi-continuous,
can be accommodated.
In the simple one dimensional case
treated here, the independent variable is
time, wavelength, or some other quantity.
The {\it data space} is the domain 
of this variable over which 
measurements were made
 -- typically an interval,
possibly interrupted by gaps during which the 
measuring system was not operating.

The measured quantity 
can be almost anything
that yields information about the target signal.
The three most common examples 
emphasized here are: 
(a) times of events
(often called time-tagged event data, or TTE),
(b) counts of events in time bins,
and
(c) measurements of 
a quasi-continuous observable
at a sequence of points in time.
For the first two cases the signal of interest 
is the \emph{event rate}, 
proportional to the probability distribution 
regulating events which occur at discrete 
times due to the nature of the 
astrophysical process and/or
the way it is recorded.
We call case (c) \emph{point measurements},
not to be confused with \emph{point data}
(also called event data).
These modes have much in common, 
as they all comprise measurements
that can be at any time; 
what differentiates them is their statistics,
roughly speaking Bernoulli, Poisson, and 
Gaussian (or perhaps some other)
respectively.

The archetypal example of (a) is 
light collected by a telescope and
recorded as a set of detection times
of individual photons
to study source variability.
Case (b) is similar, but with the
events collected into bins -- which
do not have to be equal or evenly spaced.
Case (c) is common when 
photons are not detected individually,
such as in radio flux measurements.
In all cases it is useful 
to represent the measurements with 
\emph{data cells}, typically one for each measurement
(see \S \ref{data_cells}).
In principle mixtures of cells 
from different data types 
can be handled, 
as described in the next section.

\subsection{Mixed Data Modes}
\label{mixed}

Our algorithm can analyze 
mixtures of any data types within a single time series.
For example the data stream could consist
of arbitrary combinations of cells of the
three types defined above -- 
measured values, counts in bins and events --
with or without overlap in time 
among the various data modes.
In regions of overlap the block representation
would be based on the combined data;
otherwise it would represent 
block structures supported by the corresponding
individual data modes.
In such applications 
the cost function must 
refer to a common signal amplitude parameter,
possibly including normalization 
factors to account for 
differences in the measurement processes.

A related concept is that of 
\emph{multivariate time series},
usually referring to 
concurrent observations from
different telescopes.
The distinction between this concept 
and mixed data modes is largely semantic.
Hence we leave implimentation details to 
\S\ref{multivariate}.



\subsection{Gaps}
\label{gaps}

In some cases there are 
subintervals over which no data can be obtained. 
For example 
there may be 
random interruptions such as detector malfunction 
at unpredictable but known periods of time,
or regular interruptions as 
the Earth periodically blocks the view 
of an object from an orbiting space observatory.
(Of course this case is very different from intervals 
in which no events happened to be detected,
due to low event rate, 
or in which one simply did not
happen to make point measurements.

Such data gaps have a nearly 
invisible affect on the algorithm,
fundamentally due to the fact that 
it operates locally in the time domain.
For event data all that matters is 
the \emph{live time} during the block,
\emph{i.e.} the time over which data could have been registered.
Other than correcting the total time span
of any putative block containing data gaps 
by subtracting the corresponding 
\emph{dead time},
gaps can be handled by ignoring them.
Operationally one simply treats 
the data right after a gap 
as immediately following 
the data right before it 
(and not delayed by the length of the gap).
Think of this as squeezing the interval 
to eliminate the gaps, carrying out 
the analysis as if no gaps are present,
and then un-doing the squeezing by 
restoring the original times.
This procedure is valid 
because event independence
means that the fitness of a block
depends on only 
its total live-time 
and the events within it.

For event data this squeezing can be 
implemented by 
subtracting from each event time
the sum of the lengths of all the 
preceding gaps.
One small detail concerns the points 
just before and just after a gap.
One might think 
their time intervals should be 
computed relative to the gap edges.  
But it follows from the nature of 
independent events (Appendix B, \S\ref{appendix_b})
that they can be computed as though 
the gap did not exist.

The only other subtlety lies in interpreting the model
in and around gaps. 
There are two possibilities: a given gap 
(a) may lie completely within a block
or (b) it may separate two blocks.
Case (a) can be taken as evidence that 
the event rates before and after the gap
are deemed the same within statistical fluctuations.
Case (b) on the other hand implies that
the event rate did change significantly.

Of course the gaps must be restored for display and
other post-processing analysis.
Think of this as un-squeezing the data 
so that all blocks appear at their correct locations in time.
Keeping in mind 
that there is no direct information about
what happened during unobserved 
intervals,
plots should probably include 
some indication that rates within gaps 
are unknown or uncertain, such as
by use of dotted lines or shading in
the gap for case (a) or leaving the
gap interval completely blank in case (b).

For the case of point measurements
the situation is different.  
In one sense there are no gaps at all,
and in another sense the entire observation 
interval consists of many gaps separating
tiny intervals over which the measurements
were actually made.  
One is hard-pressed to 
make a statistical distinction between 
various reasons why there is not a measurement
at a given time -- \emph{e.g.} detector and weather
problems, or simply a choice as to how to allocate
observing time (a choice that may even 
depend on the results of
analyzing previous data).
Basically the blocks in this case 
represent intervals where 
whatever measurements were made in the interval
are consistent with a signal 
that is constant over that interval.

Note that things would be different 
if one wanted to define 
a fitness function dependent
on the total length of the
block, not just its live time.
This would arise for example 
if a prior on the block length were imposed.
Such possibilities will not be discussed here.

\subsection{Exposure Variations}
\label{exposure_variations}

In some applications the effective 
instrument response is not constant.
The measurements then reflect 
true source variations 
modified by changes in overall 
throughput of the detection system.
We use the term \emph{exposure} 
for any such effect on the
detected signal -- \emph{e.g.} detector efficiency, 
telescope effective area, 
beam pattern and point spread 
function effects.
Exposure can be quantified by the ratio of 
the expected signal
with and without any such effects.
It may be calculable from properties
of the observing system, determined
after the fact through some sort of
calibration procedure,
or a combination of the two.
Here we assume that this ratio is known
and expressed
as a number $e_{n}$, typically with $0 \le e_{n} \le 1$,
for data cell $n$.

The adjustment for exposure is simple,
namely change the parameter 
representing the observed signal amplitude 
in the likelihood 
to what it would have been if the
exposure had been unity.
First compute the exposure $e_{n}$ for
data cell $n$.
Then increase by the factor $1 / e_{n}$ 
whatever quantity in the data cell 
represents the measured signal intensity.
Specifically, for time-tagged event data
this parameter is the reciprocal
of the interval of the corresponding data cell:
$1 / \Delta t_{n}$
(see eq. (\ref{interval_1})),
which is then replaced with
$1 / ( e_{n} \Delta t_{n} )$.
For bin counts the bin size can be 
multiplied by $e_{n}$ or equivalently 
the count by $1 / e_{n}$.
For point measurements multiply the amplitude
measurement by $1 / e_{n}$
(and adjust any observational error parameters accordingly).
In all cases the goal is 
to represent the data
as closely as possible to 
what it would have been if the exposure 
had been constant.
Of course this restoration is not exact in individual cases, but is correct on average.

For TTE data the fact that interval $\Delta t_{n}$ as
we define it in
eq. (\ref{interval_1})
depends on the times of two different
events  (just previous to
and just after the one under consideration) 
may seem 
to pose a problem.
The exposures of these events will in general be 
different,
so what value do we use for 
the given event?
The comforting answer is that the only relevant
exposure is that for the given event itself.
For consider the interval from the previous
to the current time, namely 
$t_{n} - t_{n-1}$.
Here $t_{n-1}$ is regarded as simply a fiducial
time and the distribution of this interval
is given by 
eq. (\ref{poisson_intervals})
with $\lambda$ the
true rate adjusted by the exposure
for event $n$,
by the principle described in \S\ref{poisson_nature}
just after this equation.
Similarly by a time-reversal invariance 
argument the distribution of the
interval to the subsequent event,
namely $t_{n+1} - t_{n}$,
also depends on only the same quantity.
In summary event independence 
(Appendix C, \S\ref{appendix_c})
yields the somewhat counterintuitive fact that 
the probability distribution of
$\Delta t_{n} = ( t_{n+1} - t_{n-1} ) / 2$
of the interval surrounding event $n$
depends on only the effective event
rate for event $n$.



\subsection{Prior for the Number of Blocks}
\label{ncp_prior}

Earlier work \cite{scargle_v} did not assign an explicit prior probability
distribution for the number of blocks, \emph{i.e.}  the
parameter $N_{blocks}$. This omission amounts to using a flat prior,
but in many contexts it is unreasonable to assign the same prior
probability to all values. 
In particular, in most settings 
it is much more likely {\it a priori}  that 
$N_{blocks} << N$ than that 
$N_{blocks} \approx N$. 
For this reason it is desirable to impose a
prior that assigns smaller probability to a large number of blocks,
and we adopt this {\it geometric prior} \cite{coram}:
\begin{equation}
P( N_{blocks} ) =
  P_{0} \gamma ^{ N_{blocks} }
\label{prior}
\end{equation}
\noindent for $0 \le N_{blocks} \le N$, and zero otherwise since
$N_{blocks}$ cannot be negative or larger than the number of data
cells. The normalization constant $P_{0}$
is easily obtained, giving
\begin{equation}\label{full_prior}
    P( N_{blocks} ) =
  {  1 - \gamma  \over 
  1 - \gamma ^{N+1} } \ \gamma ^{ N_{blocks} }
\end{equation}
\noindent 
Through this prior the parameter $\gamma$
influences the number of blocks in the optimal
representation -- a number of some importance since it affects the
visual appearance of the representation and to a lesser extent the
values of quantities derived from it. 
This form for the distribution 
dictates that finding $k+1$ blocks is less likely by 
the constant factor $\gamma$ than is finding $k$ blocks.  
In almost all applications $\gamma$ 
will be chosen  $<1$ to express 
that a smaller number of blocks 
is \emph{a priori} more likely
than a larger number.

In principle the choice of a prior 
and the values of its parameters 
expresses one's prior knowledge
in a specific problem.
The convenient geometric prior adopted
here has proven to be generic and flexible,
and its parameterization is 
simple and straightforward.
These properties are 
appropriate for a generic analysis tool
meant for a wide variety of applications.
One can think of selecting $\gamma$ as a 
simple way of adjusting the amount of 
structure in the block representation
of the signal.
It is specifically not a smoothing 
parameter but is analogous to one.

The expected number of blocks
follows from eq. (\ref{prior})
\begin{equation}\label{expect}
    <N_{blocks}>  \ = P_{0} \sum_{N_{blocks} = 0}^{N}  N_{blocks} \gamma ^{ N_{blocks} } =
     { N  \gamma  ^{N+1} +1  \over
   \gamma ^{N+1} - 1  }  + { 1 \over    {1 - \gamma  } }
\end{equation}
\noindent
Note that the actual number of blocks is 
a discontinuous, monotonic function of $\gamma$,
and because its jumps can be $>1$ 
it is not generally possible to force
a prescribed number by adjusting 
this parameter.

The above prior is not the only one possible
and different forms may be useful in some applications.
But Eq. (\ref{prior}) is very convenient to implement, 
since with the fitness equal to the
log of the posterior, one only needs to subtract the constant $-log \
\gamma$ 
(called \verb+ncp_prior+ in the MatLab code
and in the discussion of computational issues
below)
from the fitness of each block. 
Determining the value 
to use in applications is 
discussed in \S\ref{determine_prior} below.


\section{Optimum Segmentation of Data on an Interval}
\label{opt_segmentation}

Piecewise constant modeling
of sequential measurements
on a time interval  \interval \
is most conveniently implemented
by seeking 
an \emph{optimal partition} of  
the ordered set of data cells within 
\interval.
In this special case of segmentation 
the segments cover the whole set 
with no overlap between them (Appendix B).
Segmentations with overlap are possible,
for example in the case of correlated measurements,
but are not considered here.
One can envision our quest for 
the optimal segmentation as
nothing more than finding the 
best step-function, 
or piecewise constant model,
fit to the data -- defined by maximizing 
a specific fitness measure
as detailed 
in \S \ref{fitness_general}.

We introduce 
our algorithm in a somewhat abstract setting 
because the formalism developed here 
applies to
other data analysis problems beyond time series analysis. 
It implements Bayesian Blocks or other 1D segmentation
ideas for any model fitness function that satisfies a simple
additivity condition. 
It improves the previous approximate
segmentation algorithm \cite{scargle_v} by achieving 
an exact, rigorous solution of the multiple change-point problem, 
guaranteed to be a global optimum, not just a local one.

The rest of this 
section describes how the model is 
structured 
for effective solution of this problem,
while details on 
the quantity to be optimized 
are deferred to the next section.

\subsection{Partitions}
\label{partitions}

{\it Partitions} of a time interval \interval \ are 
simply collections of non-overlapping blocks 
(defined below in \S \ref{blocks}),
defined by specifying 
the number of its  blocks
and the block edges:
\begin{equation}
\mbox{{\MyScript P}}(I) \equiv \{ N_{blocks}; \ n_{k}, \ \ k = 1, 2,
3, \dots N_{blocks} \} \ . \label{partition}
\end{equation}
\noindent
where the $n_{k}$ 
are indices of the data cells (\S \ref{data_cells})
defining times called \emph{change-points} 
(see \S \ref{change-points}).

Appendix B gives a few mathematical details about
partitions, including justification of the restriction
of the change-points to coincide with data points 
and the result that the number of possible partitions
(i.e. the number of ways $N$ cells 
can be arranged in blocks) 
is $2^{N}$.
This number is exponentially large, 
rendering an explicit exhaustive search of partition space
utterly impossible for all but very small $N$.
{\bf Our algorithm implicitly performs
a complete search of this space in time of order $N^{2}$},
and is practical even for $N \sim 1,000,000$,
for which approximately $10^{300,000}$ partitions are possible.
The beauty of the algorithm is that it 
finds the optimum among all partitions without
an exhaustive explicit search, 
which is obviously impossible for almost any 
value of $N$ arising in practice.


\subsection{Data Cells}
\label{data_cells}

For input to the algorithm
the measurements are represented 
with 
\emph{data cells}.
For the most part there is one cell 
for each measurement,
although in the case of TTE data 
two or more events with identical 
time-tags may be 
combined into a single cell.
A convenient data structure 
is an array containing the cells 
ordered by the measurement times.

Specification of the contents of the cells
must meet two requirements.
First they must include time information 
allowing determination of 
which cells lie in a block
given its start and stop times.
Post-processing steps such as plotting 
the blocks may in addition use 
the actual times, either absolute or
relative to a specified origin.

The other requirement is that 
the fitness of a block 
can be computed from the contents of 
all the cells in it (\S\ref{fitness_general}, \S\ref{fitness_functions}).
For the three standard
cases the relevant data are roughly speaking:
(a) intervals between events 
(\S\ref{event_data}), 
(b) bin sizes, locations and counts 
(\S\ref{binned_event_data}),
and (c) 
measured values 
augmented by a quantifier of 
measurement uncertainty
(\S\ref{point_measurements}).
These same quantities enable 
construction of the resulting step function
for post-processing steps such as 
computing signal parameters.


\subsection{Blocks of Cells}
\label{blocks}

A {\it block} is any set of consecutive cells, 
either an element of the optimal representation 
or a candidate for it.
Each block represents a subinterval
(within the full range of observation times)
over which the amplitude of the signal 
can be estimated from 
the contents of its cells (\S\ref{data_cells}).
A block can be as small as one cell or
as large as all of the cells.

Our time series model consists of a set of blocks
partitioning the observations.
All model parameters are constant 
within each block but undergo discrete jumps
at the change-points (\S \ref{change-points}) marking the edges of the
blocks. 
The model is visualized 
by plotting rectangles spanning 
the intervals covered by the blocks,
each with height equal to the signal 
intensity averaged over the interval.
The concept of {\it fitness of a block} is fundamental to
everything else in this paper.
As we will see in the next section
the fitness of a partition 
is the sum of the fitnesses of the blocks
comprising it.


\subsection{Fitness of Blocks and Partitions}
\label{fitness_general}

Since the goal is to represent the data as well as possible within a
given class of models, we maximize a
quantity measuring the fitness of models in the given class, here
the class of all piecewise constant models. 
Alternatively, one can
minimize an error measure.
Both operations are called {\it optimization}.}
The algorithm relies on 
the fitness being block-additive, \emph{i.e.}
\begin{equation}
F[\mbox{\MyScript P}(\mbox{\interval})] = \sum_{k=1}^{N_{blocks}} f( B_{k} ) \ \ ,
\label{additivity}
\end{equation}
\noindent where $F[\mbox{\MyScript P}(\mbox{\interval})]$ 
is the total fitness
of the partition {\MyScript P} of interval \interval, and
$f( B_{k} )$ is the fitness of block $k$. 
The latter can be any convenient measure 
of how well a constant signal represents the data within the block.
Typically additivity 
results from independence of the observational errors.
We here ignore the possibility 
of correlated errors, which could make 
the fitness of one block depend on that of its neighbors.
Remember correlation of observational errors
is quite separate from correlations in the signal itself.

All model parameters are marginalized except 
the $n_{k}$ specifying block edges.
Then the total fitness depends on only these
remaining parameters -- \emph{i.e.} on the 
detailed specification of the partition by 
indicating which cells fall in each of its blocks.
The best model is found by maximizing $F$
over all possible such partitions. 

\subsection{Change-points}
\label{change-points}

In the time series literature 
a point at which a statistical model undergoes an abrupt
transition, 
by one or more of its parameters 
jumping instantaneously to a
new value, is called 
a \emph{change-point}. 
This is exactly what happens at
the edges of the blocks in our model.
In principle change-points can be at 
arbitrary times.
However,
following the data cell representation and 
without any essential loss of generality 
they can be restricted to 
coincide with a data point
(Appendix B; \S\ref{appendix_b}).

A few comments on 
notation are in order.
We take blocks to 
start at the data cell identified by the algorithm as a change-point
and to 
end at the cell previous to the subsequent change-point.
A slight variation of this convention is discussed below
in \S \ref{empty_blocks} in connection with 
allowing the possibility of empty blocks
in the context of event data.
One might adopt 
other conventions, 
such as 
apportioning the change-point data cell
to both blocks,
but we do not do so here.
Even though the first data cell in the time series 
always starts the first block, our convention is
that it is not considered a change-point.
In the code
presented here the first change-point 
marks the start of the second block.
For $k>1$ the $k$-th block 
starts at index $n_{k-1}$ and 
ends at $n_{k} -1$. 
The first block always starts with the
very first data cell.
The last block always terminates with 
the very last data cell. 
If the last cell is a change-point, 
it defines a block consisting of 
only that one cell.
The set of change-points is empty if
the best model consists of a single
block, meaning that the time series is 
sensibly constant over the whole
observation interval.
The number of blocks is one greater
than the number of change-points.


\subsection{The Algorithm}
\label{algorithm}

We now outline the basic algorithm 
yielding the desired optimum partitions.
The details of this \emph{dynamic programming}\footnote{Bellman's
explanation (before the word
``programming'' took on its current computational 
connotation) of how he chose this name is interesting.
The Secretary of Defense at the time
``... had a pathological fear and hatred of the word, research. ...
You can imagine how he felt, then, about the term, mathematical.
... I felt I had to do something to shield ... the Air Force 
from the fact that I was really doing mathematics 
inside the RAND Corporation. ...  I was interested in planning ... 
But planning is not a good word for various reasons.
I decided therefore to use the word, Ôprogramming.Õ
I wanted to get across the idea that this was dynamic ...  it's 
impossible to use the word, dynamic, in a pejorative sense. 
Try thinking of some combination
that will possibly give it a pejorative meaning.
It's impossible. Thus, I thought dynamic programming was
a good name. It was something not even a Congressman
could object to.}
approach \cite{bellman,hubert,dreyfus} 
are in \cite{jackson}.
It follows the spirit of mathematical induction:
beginning with the first data cell,
at each step one more cell is added. 
The analysis makes use of 
results stored from all previous steps.
Remarkably the algorithm is
exact and yields the optimal 
partition of
an exponentially large space
in time of order $N^{2}$.
The iterations normally continue 
until the whole interval has been analyzed. 
However its recursive nature 
allows the algorithm to function in a \emph{trigger mode},
halting when the
first change-point is detected
(\S \ref{triggers}).

Let 
$\mbox{\MyScript P}^{opt}(R)$
denote the optimal partition of 
the first $R$ cells.
In the starting case $R=1$
the only possible partition
(one block consisting of the first cell by itself) is trivially optimal.
Now assume we have completed step $R$,
identifying the optimal partition
$\mbox{\MyScript P}^{opt}(R)$.
At this (and each previous) step
store the value of optimal fitness itself in
array \textit{\textbf{best}} 
and the location of the last change-point 
of the optimal partition 
in array \textit{\textbf{last}}.

It remains to show how to obtain 
$\mbox{\MyScript P}^{opt}(R+1)$.
For some $r$ consider the set of all partitions 
(of these first $R+1$ cells) 
whose last block starts with cell $r$ 
(and by definition ends at $R+1$).
Denote the fitness of this last block by $F(r)$.
By the subpartition result in Appendix B
the only member of this set that 
could possibly be optimal 
is that consisting of  $\mbox{\MyScript P}^{opt}(r-1)$
followed by this last block. 
By the additivity in Eq. (\ref{additivity})
the fitness of said partition is 
the sum of  $F(r)$
and the fitness of $\mbox{\MyScript P}^{opt}(r-1)$
saved 
from a previous step:
\begin{equation}\label{a_r}
A(r) = F(r) + \mbox{\Huge\{ }
\begin{array}{lll}
&0 &r = 1 \\
&\verb+best+(r-1), &r = 2, 3,  \dots , R+1 \ \ .
\end{array}
\end{equation}
\noindent
$A(1)$ is the special case where the last block
comprises the entire data array and thus no previous
fitness value is needed.
Over the indicated range of $r$
this equation expresses the fitness of
all partitions $\mbox{\MyScript P}$(R+1)
that can possibly be optimal.
Hence the value of $r$ yielding the optimal partition 
$\mbox{\MyScript P}^{opt}(R+1)$
is the easily computed value maximizing $A(r)$: 
\begin{equation}
r^{opt} = \mbox{argmax}[ A(r) ] \ .
\end{equation}
\noindent
At the end of this computation, when $R=N$,
it only remains to find the locations of
the change-points of the optimal partition.
The needed information is contained in
the array 
\textit{\textbf{last}}
in which we have
stored the index $r^{opt}$ at each step.
Using the corollary 
in Appendix B 
it is a simple matter to use the last value
in this array to determine the last change-point
in $P^{opt}(N)$,
peel off the end section of 
\textit{\textbf{last}}
corresponding to this last block,
and repeat.  That is to say, the set of values
\begin{equation}
cp_{1} = \textit{\textbf{last}}(N);  \ \ 
cp_{2} = \textit{\textbf{last}}( cp_{1} - 1); \ \ 
cp_{3} = \textit{\textbf{last}}( cp_{2} - 1); \ \ \dots 
\label{cp_find}
\end{equation}
\noindent
are the index values giving the locations of the
change-points, in reverse order.
Note that the positions of the change-points are not necessarily
fixed until the very last iteration, although in practice
it turns out that they become more or less
``frozen'' once a few succeeding change-points
have been detected.
MatLab\footnote{\ \texttrademark The Mathworks, Inc} code 
for optimal partitioning of event data 
is given in Appendix A. 

\subsection{Fixing the Parameter in the Prior Distribution for  $N_{blocks}$}
\label{determine_prior}

As mentioned in \S\ref{ncp_prior} the output of the
algorithm is dependent on value of the parameter 
$\gamma$,
characterizing the assumed prior distribution
for the number of blocks, 
eq. (\ref{prior}).
In many applications the results are
rather insensitive to the value
as long as the signal-to-noise ratio 
is even moderately large.
Nevertheless extreme values of this parameter 
give bad results 
in the form of clearly too few or too many blocks.
In any case 
one must select a value to use in applications.

\begin{figure}[htb] 
\includegraphics[width=5in,height=3in]{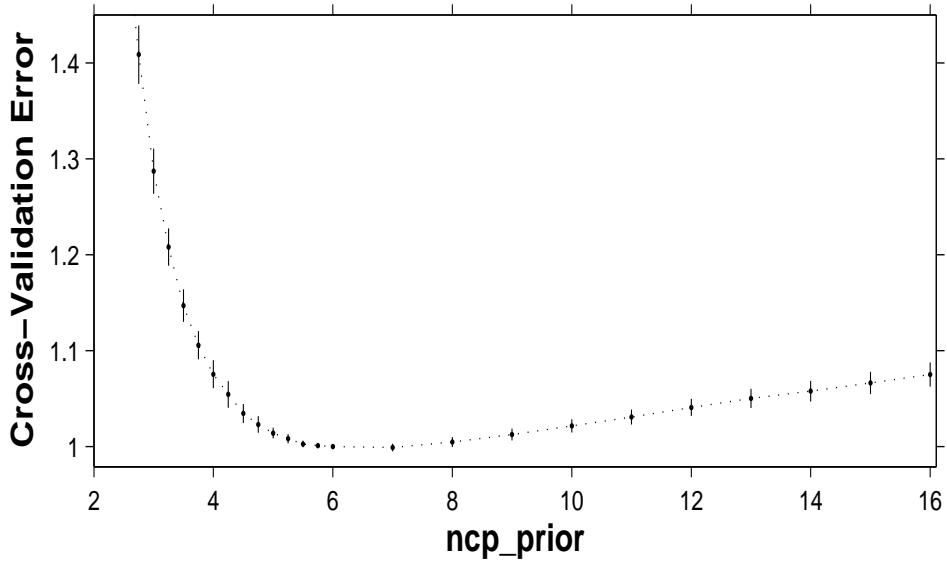} 
\caption{Cross-validation error
of BATSE TTE data (averaged over 532 GRBs, 
8 random subsamples, and time) 
for a range of values of 
-log $\gamma = $
with $3\sigma$ error bars.}
\label{batse_curve}
\end{figure}

This situation is much like that of selecting a 
smoothing parameter in various data analysis
applications, \emph{e.g.} density estimation.
In such contexts there is no perfect choice
but instead a tradeoff between bias and variance.
Here the tradeoff is between 
a conservative choice not fooled by
noise fluctuations but potentially missing real changes, 
and a liberal choice better capturing 
changes but yielding some false detections.
Several approaches have proven useful
in elucidating this tradeoff.
Merely running the algorithm 
with a few different values 
can indicate a range over which 
the block representation is reasonable
and not very sensitive to the parameter value
(\emph{cf.} Fig. \ref{batse_curve}).

The discussion of fitness functions below 
in \S\ref{fitness_functions}
gives implementation details  of 
an objective
method for calibrating  \verb+ncp_prior+
as a function of the number of data points.
It is based on 
relating this parameter 
to the  \emph{false positive} probability -- 
that is, the relative frequency with which 
the algorithm falsely reports detection of 
a change-point in data with no signal present.
It is convenient to use the complementary quantity
\begin{equation}
p_{0} \equiv 1 - \mbox{false positive probability} \ .
\label{true_positive}
\end{equation}
\noindent
This number is the frequency with which the algorithm
correctly rejects the presence of a change-point
in pure noise.
Therefore it is also the probability that a
change-point reported by the algorithm 
with this value of \verb+ncp_prior+ 
is indeed statistically significant -- 
hence we call it the  
\emph{correct detection rate} for single change-points.

The needed 
\verb+ncp_prior+-$p_{0}$ 
relationship is easily found by noting that 
the rates of correct and  incorrect responses to 
fluctuations in simulated pure noise can be controlled by 
adjusting the value of  \verb+ncp_prior+.
The procedure is:
generate a synthetic pure-noise time series;
apply the algorithm for a range of \verb+ncp_prior+;
select the smallest value 
that yields false detection frequency 
equal or less than the desired rate, such as .05.
The values of \verb+ncp_prior+
determined in this way are 
averaged over a large number of realizations
of the random data.
The result depends on only the number
of data points 
and the adopted value of   $p_{0}$:
\begin{equation}
{\mbox ncp}\_{\mbox prior} = \psi ( N, p_{0} ) \ .
\label{prior_vs_n}
\end{equation}
Results from simulations of this kind are given 
below 
for the various fitness functions in \S\S
\ref{event_data}, \ref{binned_event_data}, 
and \ref{point_measurements}.
We have no exact formulas, but rather
fits to these numerical simulations.

The above discussion is useful in  
the simple problem of deciding 
whether or not a signal 
is present above a background of noisy
observations.
In other words we have a procedure for 
assigning a value of \verb+ncp_prior+ 
that results in an acceptable frequency of 
spurious change-points, or false positives,
when searching for a single statistically significant change.  
Real-time triggering on transients
(\S\ref{triggers}) is an example of this situation,
as is any case where 
detection of a single change-point 
is the only issue in play.

But elucidating the shape of an actual detected signal
lies outside the scope of the above procedure, 
since it is based on a pure noise model.
A more general goal
is to limit the number of both false negatives and
false positives in the context of an extended signal.
The choice of the parameter value here 
depends on the nature of the signal
present and the signal-to-noise level. 
One expects that somewhat larger values of 
\verb+ncp_prior+ are necessary 
to guard against corruption of the 
estimate of the signal's shape 
due to  errors at multiple change-points.

This idea suggests 
a simple extension of the above procedure.
Assume that a value of 
$p_{0}$, the probability of correct detection
of an \emph{individual} change-point,
has been adopted and the corresponding
value of \verb+ncp_prior+ 
determined
with pure noise simulations as outlined above
and expressed in eq. (\ref{prior_vs_n}).
For a complex signal our goal is 
correct detection of  not just one, but
several change-points, say $N_{cp}$ in number.
The trick is to treat each of them 
as an independent  
detection of a single change-point 
with success rate $p_{0}$.
The probability of all $N_{cp}$ successes 
follows
from the law of compound probabilities:
\begin{equation}
p(N_{cp}) = p_{0}^{N_{cp}}  \ .
\label{multiple}
\end{equation}
\noindent
There are problems with this analysis in that 
the following are {\bf not} true:
\begin{description}
\item{(1)} Change-point detection 
in pure noise and in a signal are the same.
\item{(2)}  The detections are independent of each other.
\item{(3)} We know the value of  $N_{cp}$.
\end{description}
All of these statements would have to be true
for eq. (\ref{multiple}) to be rigorously valid.
We propose to regard the first
two as approximately true and address the
third as follows:
Decide that the probability of correctly detecting all the 
change-points should be
at least a high as some value $p_{*}$,
such as $0.95$.
Apply the algorithm using the value of 
\verb+ncp_prior+ = $\psi ( N, p_{*} )$
given by the pure noise simulation.
Use eq. (\ref{multiple}) and
the number of change-points thus found to yield
a revised value
\begin{equation}
{\mbox ncp}\_{\mbox prior} = \psi ( N, p_{*}^{1 / N_{cp} } ) \ .
\label{ncp_updata}
\end{equation}
\noindent
Stopping when 
the iteration produces 
no further modification of 
the set of change-points,
one has the
recommended value of \verb+ncp_prior+.
This ad hoc procedure is not rigorous, but it establishes
a kind of consistency and has proven useful 
in all the cases where we have tried it, 
e.g. \cite{norris1,norris2}.

Fig. \ref{batse_curve} shows another approach, based on
cross-validation of the data being analyzed
(\emph{cf.} \cite{hogg}).
This study uses the collection of 
raw TTE data at the BATSE web site\\
{\normalsize
\verb+ftp://legacy.gsfc.nasa.gov/compton/data/batse/ascii_data/batse_tte/+}.
The files for each of 532 GRBs 
contain time tags for all photons
detected for that burst.
The energy and
detector tags in the data files were not used here,
but \S\ref{ex_tte} shows an example using the former.
An ordinary 256-bin histogram of 
all photon times for each of 532 GRBs
was taken as the true signal for that burst.
Eight random subsamples 
smaller by a factor of $8$ were
analyzed with the algorithm 
using the fitness in eq. (\ref{fitness_event}). 
The average RMS error between
these block representations
(evaluated at the same 256 time points)
and the histogram is roughly flat over a 
broad range.
While this illustration with 
a relatively homogeneous data set
should obviously not be taken as universal,
the  general behavior seen here -- determination
of a broad range of nearly equally optimal 
values of \verb+ncp_prior+ -- 
is characteristic of a wide variety of situations.

\subsection{Analysis of Variance}
\label{error_analysis}

Assessment of uncertainty 
is an important part of any data analysis procedure.
The observational errors discussed throughout 
this paper are propagated by the algorithm 
to yield corresponding uncertainties in the 
block representation and its parameters.
The propagation of stochastic variability in 
the astronomical source is a separate issue,
called \emph{cosmic variance}, 
and is not discussed here.

Since the results here 
comprise a complete function
defined by a variable number of parameters, 
quantification of uncertainty is considerably 
more intricate than for a single parameter.
In particular one must specify precisely which
of the block representation's aspects 
is at issue.  Here we discuss three:
(a) the full block representation,
(b) the very presence of the change-points themselves,
and
(c) locations of change-points.

A straightforward way to deal with (a) 
is by bootstrap analysis.
As described in \cite{efron} 
for time series data
this procedure
is rather complicated
in general. However resampling of 
event data in 
the manner appropriate to the bootstrap 
is trivial.
The procedure is to 
run the algorithm on each of many 
bootstrap samples 
and evaluate the resulting block representations
at a common set of evenly spaced times.
In this way models with different numbers and locations
of change-points can be added, 
yielding means and
variances for the estimated block light curves.
The bootstrap 
variance is an indicator of light curve uncertainty.
In addition comparison of the 
bootstrap mean 
with the block representation 
from the actual data
adds information about modeling bias.
The former is rather
like a model average in the Bayesian context.
This average typically smoothes
out the discontinuous block edges 
present 
in any one representation.
In some applications the bootstrap 
mean may be more useful than the
block representation.

This method does not seem to 
be useful for studying uncertainty in the
change-points themselves, in particular their number,
presumably because the duplication of data points
due to the replacement feature of the resampling 
yields excess blocks 
(but with random locations and 
small amplitude variance,
and therefore with little effect on the mean light curve).

By (b) is meant an assessment of the
statistical significance of the identification
of a given change point. 
For a given change-point
we suggest quantification of this uncertainty 
by evaluating the ratio of the fitness functions
for the two blocks on either side of that change-point
to that of the
single block that would exist if the
change-point were not there.
The corresponding difference of the (logarithmic) 
fitness values should be adjusted by the
value of the constant parameter \verb+ncp_prior+,
for consistency with the way fitness is computed
in the algorithm.

Finally, (c) is easily addressed in an approximate 
way by fixing all but one change-point and computing
fitness as a function of the location of that change-point.
This assessment is approximate because by fixing
the other change-points because it ignores
inter-change-point dependences.  One then
converts the run of the fitness function with
change-point location into a normalized 
probability distribution, giving comprehensive
statistical information about that location (mean,
variance, confidence interval, etc.)

Sample results of all of these uncertainty measures 
in connection with 
analysis of a gamma-ray burst light curve
are shown below in \S\ref{ex_tte},
especially 
Fig. \ref{error_plot}.

\subsection{Multivariate Time Series}
\label{multi_variate}

Our algorithm's intentionally flexible data interface
not only allows processing 
a wide variety of data modes
but also facilitates joint analysis of mode combinations.
This feature allows one to obtain
the optimal block representation
of several concurrent data streams 
with arbitrary modes and sample times.
This analysis is joint in the sense that 
the change-points are constrained
to be at the same times for all the input series;
in other words the block edges 
for all of the input data series line up.
The representation is optimal for the data
as a whole but not for the individual time series.

To interpret the result of a multivariate analysis 
one can study the blocks in the different series
in two ways: 
(a) separately, but with the realization 
that the locations of their edges 
are determined by all the data;
or (b) in a combined representation.
The latter requires that there be 
a meaningful way to combine amplitudes.
For example the plot of a joint analysis of 
event and binned data could
simply display the combined event rate for each block,
perhaps adjusting for exposure differences.
For other modes, 
such as photon events and radio frequency fluxes,
a joint display would have to involve a
spectral model or some sort of relative normalization.
The example in \S \ref{multivariate}
below will help clarify these issues.

The idea extending the basic algorithm
to incorporate multiple time series
is simple.
Each datum in any mode has
a time-tag associated with it -- for example the event time,
the time of a bin center, or the time of a point
measurement.
The joint change-points are allowed to occur at any
one of these times.
Hence the times from all of the separate data streams
are collected together into a single ordered array;
the ordering means that the times -- as well
as the measurement data -- from the 
different modes are interleaved.
\begin{figure}[htb] 
\centering
\par
\hskip -0.2in
\includegraphics[width=5in,height=3in]{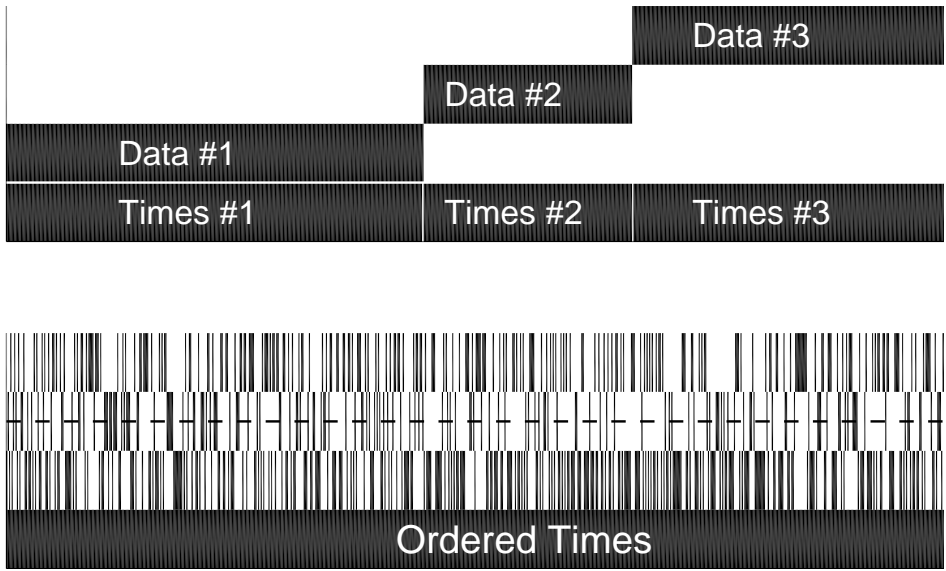} 
\caption{Cartoon depicting an example of how
three data series are first concatenated into a
matrix (top) and then redistributed by ordering 
the combined time-tags (bottom). 
The cost functions for the series can then
be computed from the data in horizontal slices
(\emph{e.g.} dashed line) and combined,
allowing the
change-points to be at any of the time tags.} 
\label{mult_pic}
\end{figure}
The cartoon in Fig. \ref{mult_pic} shows how the
individual concatenated times and data series
are placed in separate blocks in a matrix (top)
and then redistributed (bottom)
by ordering the combined times.
Then the fitness function for a given data series
can be obtained from the corresponding data slice
(e.g. the horizontal 
dashed line in the figure, for Series \#2).
The zero entries in these slices (indicated by white
space in the figure) are such that the fitness function
for data from each series 
is evaluated for only the appropriate data and mode combination.
The overall fitness is then simply the sum of those
for the several data series.
The details of this procedure are
described in the code provided in Appendix A
(\S\ref{appendix_a}).

\subsection{Comparison with Theoretical 
Optimal Detection Efficiency }

How good is the algorithm at 
extracting weak signals in noisy data?
This section gives evidence that it
achieves detection sensitivity 
closely approaching ideal theoretical limits.
The formalism
in \cite{adh} 
treats detection
of geometric objects in data spaces of
arbitrary dimension
using multiscale methods.
The one dimensional 
special case in \S II of this reference
is essentially 
equivalent to our problem
of detecting a single block in 
noisy time series.

Given $N$ measurements
normalized so that 
the observational errors $\sim N(0,\sigma)$
(normally distributed with zero mean 
and variance $\sigma ^{2}$), 
these authors show that 
the threshold for detection is
\begin{equation}\label{arias_castro}
A_{1} = \sigma \  \sqrt{ 2 \ log N } \ .
\end{equation} 
\noindent
This result is asymptotic 
(\emph{i.e.} valid in the limit of large $N$).
It is valid for a frequentist
detection strategy based on 
testing whether the 
maximum of the inner product
of the model with the data
exceeds the quantity in eq. (\ref{arias_castro})
or not.
These authors state
``In short, we can efficiently
and reliably detect intervals of
amplitude roughly 
$\sqrt{\ 2 \ \mbox{log} N  \ }$, but not smaller.''
More formally
the result is that asymptotically their test 
is powerful for signals
of amplitude greater than $A_{1}$
and powerless for weaker signals.

\begin{figure}[htb]
\includegraphics[width=5in,height=4.1in]{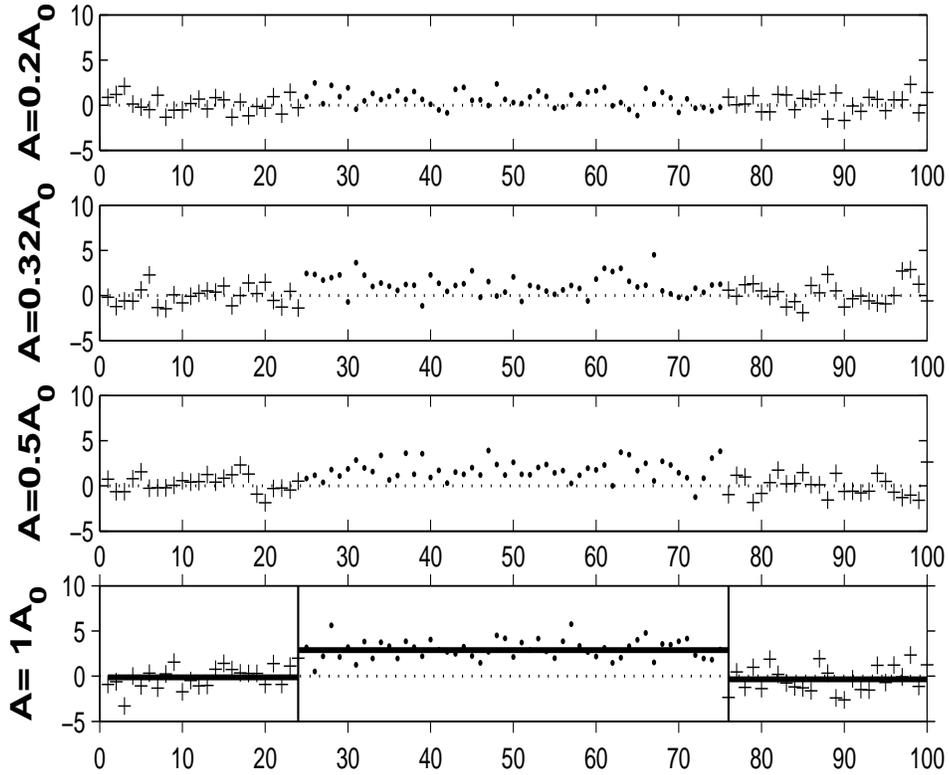} 
\caption{One hundred unit variance 
normally distributed measurements --
zero-mean (+) except for 
a block of events 25-75 (dots).
In the four panels the block amplitudes are 
0.2, .32,  .5,  and 1.0 in units of the Arias-Castro {\it et al.} threshold $\sqrt{\ 2 \  log N}$.
Thick lines show the
blocks, 
where detected, 
with thin vertical lines at the change-points.}
\label{ac_limit_1}
\end{figure}
It is of interest to see how well
our algorithm stacks up against 
these theoretical results, 
since the two analysis approaches
(matched filter test statistic vs.
Bayesian model selection)
are fundamentally different.
Consider a simulation
consisting of normally distributed 
measurements at arbitrary times in an interval.
These variates are taken to be zero-mean-normal,
except over an unknown sub-interval
where the mean is a fixed constant.
In this experiment the events are evenly spaced, but
only their order matters, so the results would be the
same for arbitrary spacing of the events.
Fig. \ref{ac_limit_1}
shows synthetic data for four
simulated realizations with different
values for this constant.
The solid line is the Bayesian
blocks representation, using the
posterior in Eq. (\ref{final_flat}).
For the small amplitudes 
in the first two panels 
no change-points are found;
these weak signals are completely missed.
In the other two panels the
signals are detected and 
approximately correctly represented -- with small
errors in the locations of the change-points.

\begin{figure}[htb]
\includegraphics[width=5in,height=4.5in]{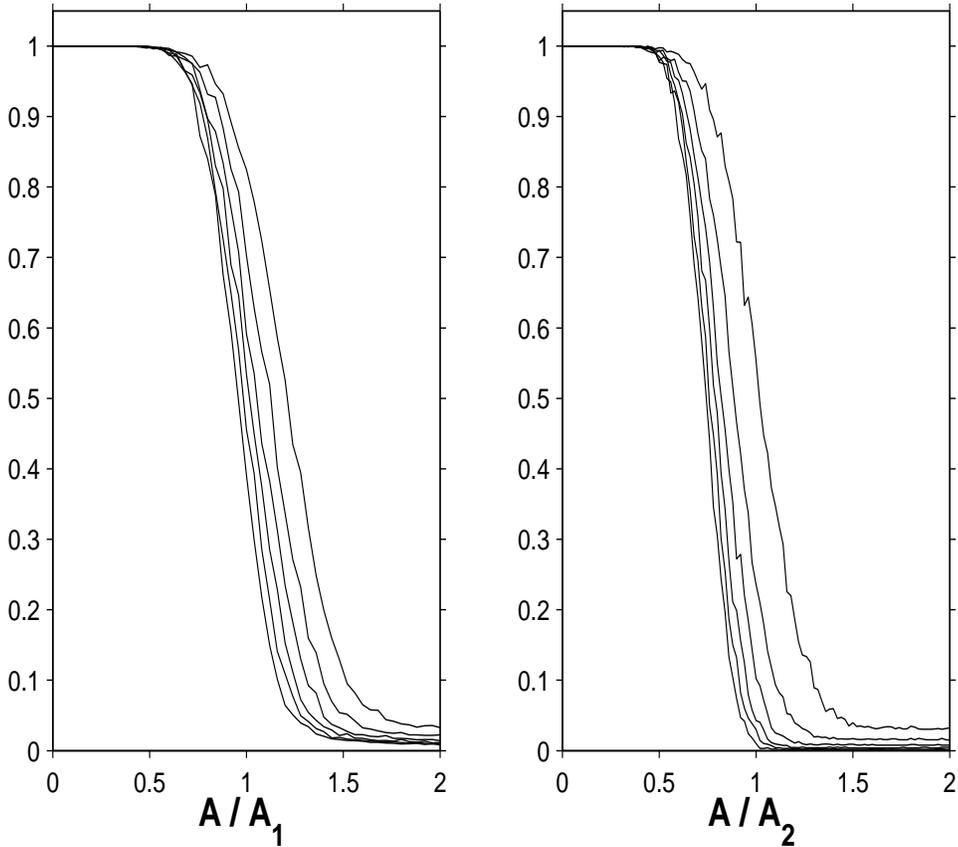} 
\caption{Error in finding a single block
vs. simulated block amplitude in units of Arias-Castro
{\it et al.}'s threshold amplitude.
The curves (from right to left) are for
N = 32, 64, 128, 256, 1024 and 2048.}
\label{ac_limit_2}
\end{figure}
Fig. \ref{ac_limit_2} reports 
some results of detection of
the same step-function
process shown in Fig. \ref{ac_limit_1},
averaged over many different
realizations of the observational error process
and for several different values of $N$.
The lines are plots of  
a simple error metric 
(combing the errors in the number 
of change-points and their locations)
as a function of the amplitude
of the test signal.
The left panel is for the case where the
number of points in the putative block is
held fixed, whereas the right panel 
is this number is taken to be proportional 
to $N$, sometimes a more realistic situation.
We have adopted the following definition
for the threshold in this case:
\begin{equation}\label{arias_castro_mod}
A_{2} = 8 \sigma \  \sqrt{ {2 \ log N \over N }} \ .
\end{equation} 
\noindent
This formula is consistent with adjusting 
the normalized width in \cite{adh}
with a factor $N$; $8$ is an
arbitrary factor for plotting.

Our method yields small errors when the 
signal amplitude is on the order
or even somewhat smaller than 
the limit stated by \cite{adh},
showing that we are indeed 
close to their theoretical limit.
The main difference here is that
our results are for specific values of
$N$ and the theoretical results are asymptotic
in $N$.

\section{Block Fitness Functions}
\label{fitness_functions}

To complete the algorithm
all that remains is to define 
the  
\emph{model fitness function} 
appropriate to a particular data mode.
By equation (\ref{additivity}) it is sufficient to 
define a 
\emph{block fitness function},
which can be any
convenient measure of 
how well a constant signal represents 
the data in the block. 
Naturally this measure will depend on 
all data in the block and not on any 
outside it.
As explained in
\S \ref{fitness_general}
it cannot depend on any model parameters
other than those specifying the locations 
of the block edges.
In practice this means that 
block height (signal amplitude) 
must somehow be
eliminated as a parameter.
This can be accomplished, for example, 
by taking block fitness to be 
the relevant likelihood 
either maximized or marginalized with respect to 
this parameter.
Either choice
yields a quantity good for comparing
alternative models, 
but not necessarily for assessing 
goodness-of-fit of a single model.
Note that these measures as such
do not satisfy the additivity condition 
Eq. (\ref{additivity}).
As long as the cell measurement errors
are independent of each other
the likelihood of a string of blocks 
is the product 
of the individual values,
but not the required sum.
But simply taking the logarithm 
yields the necessary additivity.

There is considerable
freedom in choosing fitness functions
to be used for a given type of data.
The examples described here have
proven useful in various circumstances,
but the reader is encouraged to explore
other block-additive functions
that might be more appropriate in a given 
application.
For all cases considered in this paper
the fitness function 
depends on data in the block 
through summary parameters 
called 
\emph{sufficient statistics},
capturing the statistical behavior of the data.
If these parameters 
are sums of quantities defined on the cells 
the computations are simplified; however this condition is
not essential.

Two types of factors 
in the block fitness can be ignored.
A constant factor $C$ appearing in the
likelihood \emph{for each data cell} 
yields an overall constant term
in the derived logarithmic fitness function 
for the whole time series,
namely $N \hskip .04in\mbox{log} \hskip .04in C$.
Such a term is independent of all model parameters
and therefore irrelevant for the model comparison 
in the optimization algorithm.  
In addition, while a term in the block fitness 
that has the same value  \emph{for each block}
does affect total model fitness, 
it contributes a term proportional to the number of blocks,
and which therefore can be 
absorbed into the parameter derived from the
prior on the number of blocks (\emph{cf}. \S \ref{ncp_prior}).

Many of the data modes 
discussed in the following subsections were
operative in 
the Burst and Transient Source Experiment (BATSE) experiment
on the NASA Compton Gamma Ray Observatory (GRO), the Swift Gamma-Ray Burst Mission, 
the Fermi Gamma Ray Space Telescope, 
and many x-ray and other high-energy
observatories.
They are also relevant in a wide range of other applications.

In the rest of this section
we exhibit expressions that serve
as practical and reliable fitness functions
for the three most common data modes:
event data, binned data, 
and point measurements with normal
errors.
Some refinements of this discussion
and some other less common data
modes are discussed
in Appendix C, \S\ref{appendix_c}.

\subsection{Event Data}
\label{event_data}

For series of
times of discrete events 
it is natural to associate one data cell
(\S \ref{data_cells}) with each event.
The following derivation of the appropriate 
block fitness will elucidate exactly what 
information the cells must contain to allow 
evaluation of the fitness for the full 
multi-block model.

In practice the event times 
are integer multiples of some small unit
(\S\ref{fitness_events_alt})
but it is often convenient
to treat them as real numbers 
on a continuum. 
For example the fitness function 
is easily obtained 
starting with the unbinned likelihood
known 
as the Cash statistic (\cite{cash}; a thorough discussion is in
\cite{tompkins}).
If $M( t, \theta )$ is a model 
of the time dependence of a signal
the unbinned log-likelihood is 
\begin{equation}
\label{cash}
{\mbox log} L( \theta ) = \sum_{n} 
{\mbox log} M( t_{n}, \theta ) \ - \int M( t, \theta ) dt \ ,
\end{equation}
\noindent
where 
the sum is over the events
and $\theta$ represents the model parameters.
The integral is over the observation interval and 
is the expected number of
events under the model.
Our block model is 
constant with a single parameter, 
$M(t,\lambda) = \lambda$,
so for block $k$ 
\begin{equation}
\label{cash_constant}
{\mbox log} L^{(k)}(\lambda)  = N^{(k)} {\mbox log} \lambda
 \ - \lambda T^{(k)} \ ,
\end{equation}
\noindent
where 
$N^{(k)}$ is the number of events in block $k$ 
and
$T^{(k)}$ is the length of the block.
The maximum of this likelihood is at $\lambda = N^{(k)} / T^{(k)}$,
yielding 
\begin{equation}\fbox{$ \ 
\mbox{log} \ L^{ (k) }_{max} + N^{(k)}  = 
 N^{(k)} ( \  \mbox{log} N^{(k)} - \mbox{log}  T^{(k)} ) 
 \ 
 $} \ .
 \label{fitness_event}
 \end{equation}
 \noindent
The term $N^{(k)}$ is taken to the 
left side because its sum over the blocks is  a 
constant ($N$, the total number of events)
that is model-independent and therefore irrelevant.
Moreover note that changing the units of time,
say by a scale factor $\alpha$, changes
the log-likelihood 
by $-N^{(k)}\ {\mbox log} ( \alpha )$,
irrelevant for the same reason.
This felicitous property
holds for other maximum likelihood fitness functions
and removes what would otherwise be
a parameter of the optimization.
This effective scale invariance 
and the simplicity of 
eq. (\ref{fitness_event}) make its block sum
the fitness function of choice to find the 
optimum block representation of 
event data.
A possible exception is the case where 
detection of more than one
event at a given time is not possible, 
e.g. due to detector,  \emph{deadtime},
in which case the fitness function in Appendix C, 
\S \ref{event_mode_2}
may be more appropriate.

It is now obvious what information 
a cell must contain to allow evaluation
of the sufficient statistics 
$N^{(k)}$ and $T^{(k)}$
by summing two quantities over the cells in a block.
First it must contain 
the number of events in the cell.
(This is typically one, but can be more 
depending on how duplicate
time tags are handled; see the
code section in Appendix A, \S\ref{appendix_a},
dealing with duplicate time-tags, or ones that are
so close that it makes sense to treat them as identical).
Second, it must contain the interval
\begin{equation}
\Delta t_{n} = ( t_{n+1} - t_{n-1} ) / 2  \ , 
\label{interval_1}
\end{equation}
\noindent
representing the contribution of cell $n$
to the length of the block.
This interval
contains all times closer to 
event $n$ than to any other.
It is defined by the midpoints between
successive events, and generalizes to data spaces of any
dimension, where it is called the {\it Voronoi tessellation} of the
data points, \cite{spatial_tessellations,scargle_2,scargle_4}).
Because $1 / \Delta t_{n}$ can be regarded 
as an estimate of the local event rate at time $t_{n}$,
it is natural to visualize the corresponding data cell
as the unit-area rectangle of width $\Delta t_{n}$
and height $1 / \Delta t_{n}$.
These ideas lead to the comment in 
\S\ref{exposure_variations} that the 
event-by-event adjustment
for exposure can be implemented by 
shrinking $\Delta t_{n}$  by the exposure
factor $e_{n}$.

It is interesting to note that 
the actual locations of the (independent)
events
within their block do not matter. 
The fitness function
depends on only the number of 
events in the block, 
not their locations or the intervals between them.
This result flows directly from the
nature of the underlying independently distributed,
or Poisson, process (see Appendix B, \S\ref{appendix_b}). 

We conclude this section with evaluation of the
calibration of \verb+ncp_prior+ 
from simulations of signal-free observational
noise as described in \S\ref{determine_prior}.
The results of extensive simulations 
for a range of 
values of $N$ and the adopted false positive
rate $p_{0}$ introduced in Eq. (\ref{true_positive})
were found to be well fit with the formula
\begin{equation}
\mbox{ncp}\_\mbox{prior} = 4 - 
73.53  p_{0} N^{-.478}  
\label{ncp_fit_event}
\end{equation}
\noindent
For example, with $p_{0} = .01$ and $N=1,000$
this formula gives \verb+ncp_prior+ = 3.97.
    
\subsection{Binned Event Data}
\label{binned_event_data}

The expected count in a bin is the product 
$\lambda eW$ of the
true event rate $\lambda$ at the detector, 
a dimensionless 
exposure factor  $e$ (\S\ref{exposure_variations}),
and
the width of the bin $W$.
Therefore the  likelihood for bin $n$ is given by 
the Poisson distribution
\begin{equation}
L_{n} = {(\lambda e_{n} W_{n} )^{N_{n}} e ^{ - \lambda e_{n}W_{n} }
 \over N_{n}! } \ ,
\label{bin_likelihood}
\end{equation}
\noindent 
where 
$N_{n}$ is the number of events in bin $n$,
$\lambda$ is the actual event rate in counts per unit time, 
$e_{n}$ is the exposure averaged over the bin,
and
$W_{n}$ is the bin width in time units.
Defining
\emph{bin efficiency}  as $w_{n} \equiv e_{n}W_{n}$,
the likelihood for block $k$ is the product of the likelihoods of
all its bins:
\begin{equation}
L^{(k)} = \prod_{n=1}^{M^{(k)}} L_{n} = \lambda^{N^{(k)}} e ^{ -
\lambda w^{(k)} }  .
\end{equation}
\noindent Here $M^{(k)}$ is the number of bins in block $k$,
\begin{equation}
w^{(k)} = \sum_{n=1}^{M^{(k)}} w_{n}
\end{equation}
\noindent is the sum of the bin efficiencies in the block, and
\begin{equation}
N^{(k)} = \sum_{n=1}^{M^{(k)}} N_{n}
\end{equation}
\noindent is the total event count in the block.  
The factor $(e_{n} W_{n} )^{N_{n}} / N_{n}! $ 
has been discarded because its product
over all the bins in all the blocks 
is a constant (depending on the data only)
and therefore irrelevant to model fitness.
The log-likelihood is
\begin{equation}
\mbox{log} L^{(k)} = 
{N^{(k)}} \mbox{log}  \lambda  -
\lambda w^{(k)}  \ ,
\end{equation}
\noindent
identical to eq. (\ref{cash_constant})
with $w^{(k)}$ playing the role of $T^{(k)}$,
a natural association since it is an effective block duration.
Moreover in retrospect it is understandable that
unbinned and binned event data
have the same fitness function,
especially in view of the
analysis in \S\ref{fitness_events_alt}
where ticks are allowed to contain more
than one event and are thus equivalent to bins.
In addition the way variable exposure 
is treated here could just as well have 
been applied to event data in the previous section.
Note that in all of the above 
the bins are not assumed to be equal or contiguous -- there can
be arbitrary gaps between them (\S\ref{gaps}).

\begin{figure}[htb] 
\centering
\hspace{6.9in}
\par
\hskip -0.2in
\includegraphics[width=5in,height=4in]{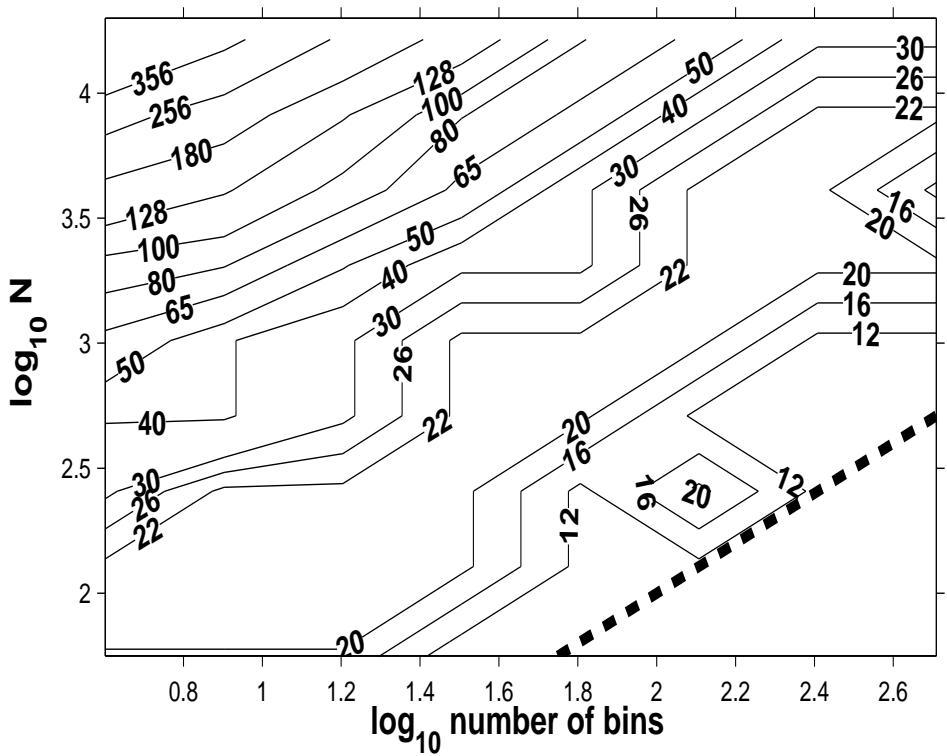} 
\caption{Simulation study, based on the false positive
rate of 0.05,  to determine  
$\mbox{ncp}\_\mbox{prior}$ = -log($\gamma$) 
for binned data. Contours of this parameter are shown as
a function of the number of bins and number of data
points (logarithmic x- and y- axes, respectively).
The heavy dashed line indicates 
the undesirable region where the numbers of
bins and data points are equal.} 
\label{calibrate_bins}
\end{figure}
We now turn to the determination of 
$\mbox{ncp}\_\mbox{prior}$ for binned data. 
Figure \ref{calibrate_bins}
is a contour plot of the values of this parameter
based on a simulation study with bins
containing independently distributed events.
These contours are very insensitive to the 
false positive rate, which was taken as $.05$ in this figure.

\subsection{Point Measurements}
\label{point_measurements}

A common experimental scenario is to
measure a signal $s(t)$
at a sequence of times
$t_{n}, n = 1, 2, \dots , N$
in order to characterize 
its time dependence.
Inevitable corruption due to observational errors
is frequently countered by smoothing the data
and/or fitting a model.
As with the other data modes 
Bayesian Blocks is a different approach to this issue,
making use of knowledge of the 
observational error distribution
and
avoiding the information loss entailed by smoothing.
In our treatment the set of observation times $t_{n}$,
collectively known as the \emph{sampling}, 
can be anything -- evenly
spaced points or otherwise.
Furthermore we explicitly assume that the 
measurements at these times are independent of each other,
which is to say the errors of observation are statistically independent.

Typically these errors
are random and additive, so that
the observed time series can be modeled as 
\begin{equation}
x_{n} \equiv 
x(t_{n}) = s(t_{n}) + z_{n} \ \ n = 1, 2, \dots N \ .
\label{obs_err}
\end{equation}
\noindent
The observational error 
$z_{n}$, at time $t_{n}$,
is known only through its statistical distribution. 
Consider the 
case
where the errors are taken 
to obey a normal 
probability distribution
with zero mean and given variance:
\begin{equation}
P(z_{n} ) dz_{n}  = {1 \over \sigma_{n} \sqrt{ 2 \pi }  } \ \
e^{ - {1 \over 2} ({ z_{n} \over \sigma_{n} })^2 } dz_{n} \ .
\label{obs_err_dist}
\end{equation}
\noindent

Using eqs. (\ref{obs_err}) and (\ref{obs_err_dist}) 
if the model signal is the constant $s = \lambda$
the likelihood of
measurement $n$ is
\begin{equation}
L_{n} = {1 \over \sigma_{n} \sqrt{ 2 \pi }  } \ \ e^{ - {1 \over 2}
({  x_{n} - \lambda \over \sigma_{n} })^2 } \ .
\label{likelihood}
\end{equation}
\noindent 
Since we assume independence of the measurements 
the block $k$ likelihood is
\begin{equation}
L^{(k)} = \prod_{n} L_{n}   =
{( 2 \pi )^{-{N_{k} \over 2}}  \over  \prod_{m} \sigma_{m}}  \ \ 
e^{ - {1 \over 2} \sum_{n}
 ({
x_{n} - \lambda \over \sigma_{n} })^2 } \ .
\label{block_like}
\end{equation}
\noindent 
Both the products and sum are over 
those values of the index such that $t$ lies in block $k$.
The quantities multiplying the exponentials in 
both the above equations 
are irrelevant
because they contribute
an overall constant factor
to the total likelihood.

We now derive 
the maximum likelihood
fitness function for 
this data mode
(with other forms based on different
priors relegated to Appendix C,
\S\S \ref{point_1}, \ref{point_2}, \ref{point_3} and \ref{point_4}).
The quantities
\begin{equation}
a_{k} = {1 \over 2} \sum_{n} {1 \over \sigma_{n}^{2}}
\label{aa}
\end{equation}

\noindent
\begin{equation}
b_{k} = - \sum_{n} {x_{n} \over \sigma_{n}^{2}}
\label{bb}
\end{equation}

\noindent
\begin{equation}
c_{k} = {1 \over 2} \sum_{n} {x_{n}^{2} \over \sigma_{n}^{2}}
\label{cc}
\end{equation}
\noindent
appear in all versions of these
fitness functions;
the first two are 
sufficient statistics.

As usual we need to remove 
the dependence of eq.\ (\ref{block_like})
on the parameter $\lambda$, and
here we accomplish this result by 
finding the value of $\lambda$
which maximizes the block likelihood,
that is 
by maximizing 
\begin{equation}
-{1 \over 2} \sum_{n} ({  x_{n} - \lambda \over
\sigma_{n} })^2 \ .
\end{equation}
This is easily found to be
\begin{eqnarray}
\label{lambda_max_1}
&\lambda_{max} &=  \sum_{n} {x_{n} \over \sigma_{n}^{2} } \ 
/ 
\sum_{n'} {1 \over \sigma_{n'}^{2} }    \\
&  &= -  b_{k} / 2 a_{k} 
\label{lambda_max_2}
\end{eqnarray}
\noindent 
As expected
this maximum likelihood amplitude
is just the weighted mean value 
of the observations $x_{n}$ within the block,
because defining the weights
\begin{equation}
w_{n} = { {1 \over \sigma_{n}^{2} }
\over 
\sum_{n'}( {1 \over \sigma_{n'}^{2} } )
} \ ,
\end{equation}
\noindent
yields
\begin{equation}\label{lam_max_wx}
\lambda_{max} =  \sum_{n} w_{n} x_{n} \ .
\end{equation}
\noindent 
Inserting Eq. (\ref{lambda_max_2})
into the log of Eq. (\ref{block_like}) 
with the irrelevant factors omitted
yields the corresponding 
maximum value of the log-likelihood itself:
\begin{equation}
\mbox{log} L^{(k)}_{\mbox{\small max} }
= 
- {1 \over 2} \sum_{n} (
{ x_{n}  + { b_{k} \over 2 a_{k} }
\over \sigma_{n} }  )^2 
\label{log_like_1}
\end{equation}
\noindent 
where again the sums
are over the data in block $k$.
Expanding the square 
\begin{equation}
\mbox{log} L^{(k)}_{\mbox{\small max} }
= 
- {1 \over 2} [ \ \
\sum_{n} { x_{n}^{2}  \over \sigma_{n}^{2} }  
+ {b_{k} \over a_{k} } \sum_{n}  { x_{n}  \over \sigma_{n}^{2} }  
+
{ b_{k}^{2} \over 4 a_{k}^{2}  } 
\sum_{n} { 1 \over \sigma_{n}^{2}  } \ \  ] \ ,
\label{log_like_2}
\end{equation}
\noindent 
dropping the first term
(quadratic in $x$) which also 
sums to a model-independent
constant, and using 
equations (\ref{aa}) and (\ref{bb})
we arrive at \begin{equation}\fbox{$ \ \ \ 
\mbox{log} L^{(k)}_{\mbox{\small max} }
=
b_{k}^{2} /   4 a_{k}
\label{log_like_4}
\\\ $}  \ .
\end{equation}
\noindent 
As expected each data cell must contain 
$x_{n}$ and $\sigma_{n}$
but we now see that these quantities 
enter the fitness function through 
the summands in the equations 
(\ref{aa}) and (\ref{bb}) defining $a_{k}$ and $b_{k}$
($c_{k}$ does not matter),
namely
$1 / (2 \sigma_{n}^{2})$
and
$-x_{n} / \sigma_{n}^{2}$.
The way the corresponding block summations
are implemented is
described in Appendix A \S\ref{appendix_a},
(\emph{c.f.} data mode \#3).

A few additional notes may be helpful.
In the familiar case 
in which the error variance is assumed to be
time-independent $\sigma$ can be carried as 
an overall constant and $\sigma_{n}$ 
does not have to be specified in each data cell.
The $t_{n}$
are only relevant in determining 
which cells belong in a block
and do not enter the fitness computation explicitly.
And the fitness function
in Eq. (\ref{log_like_4})
is manifestly invariant to a scale change 
in the measured quantity,
as is the alternative form 
derived in Appendix C, Eq. (\ref{final_ml}).
That is to say under the transformation
\begin{equation}
\label{trans}
x_{n} \rightarrow a x_{n}, 
\sigma_{n} \rightarrow a \sigma_{n} \ ,
\end{equation}
\noindent
corresponding for example to 
a simple change in 
the units of $x$ and $\sigma$,
the fitness does not change.

\begin{figure}[htb]
\includegraphics[width=5in,height=7.5in]{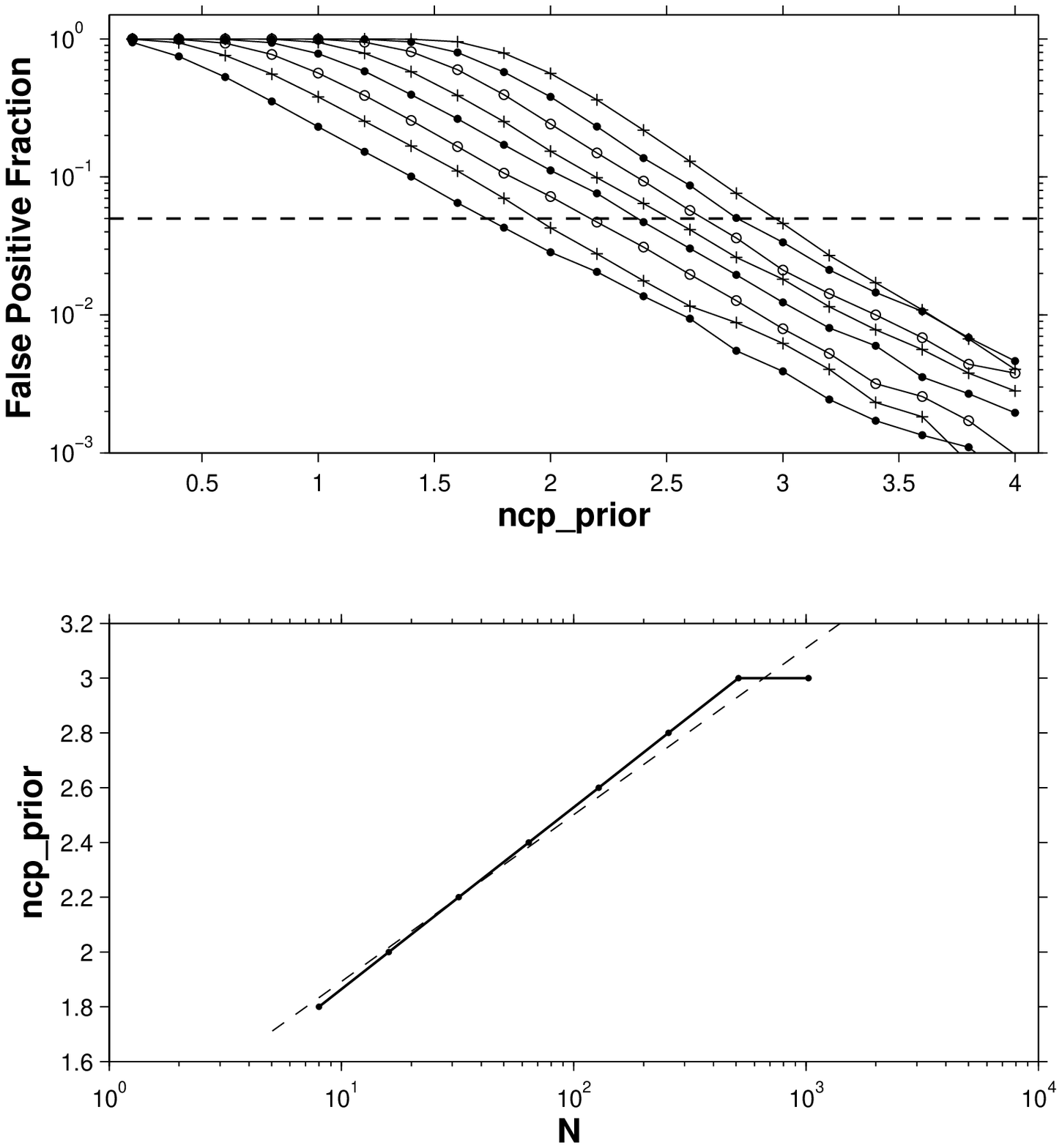} 
\caption{Simulations of point measurements
(Gaussian noise with signal-to-noise ratio of 10) to determine 
$\mbox{ncp}\_\mbox{prior}$ = -log($\gamma$).
Top: false positive fraction $p_{0}$ 
\emph{vs.} value of $\mbox{ncp}\_\mbox{prior}$
with separate curves for the values 
N = 8, 16, 32, 64, 128, 256, 512 and 1024
(left to right; alternating dots, + and circles). 
The points at which the rate 
becomes unacceptable (here .05; dashed line)
determines the recommended values of 
$\mbox{ncp}\_\mbox{prior}$ 
shown as a function of N in the bottom panel.
} 
\label{calibrate_bins_2}
\end{figure}

Figure \ref{calibrate_bins_2} exhibits a simulation 
study to calibrate $\mbox{ncp}\_\mbox{prior}$
for normally distributed point measurements.
For illustration the pure noise data simulated 
was normally distributed with a mean of 10 and unit
variance.
The left-hand panel shows how the false positive
rate is diminished as $\mbox{ncp}\_\mbox{prior}$
is increased, for the 8 values of N listed in the caption.
The horizontal line is at the adopted false positive
rate of 0.05; the points at which these curves cross 
below this line generate the curve shown in 
the bottom panel.   The linear fit in the 
latter depicts
the relation $\mbox{ncp}\_\mbox{prior}
= 1.32 + 0.577 \ \mbox{log}_{10}( N )$.
This relation is insensitive to the signal-to-noise 
ratio in the simulations.
\clearpage

\section{Examples}
\label{examples}

The following subsections present 
illustrative examples with sample data sets,
demonstrating 
block representation for TTE data,
multivariate time series,
triggering,
the empty block problem for TTE data,
and data on the circle.

\subsection{BATSE Gamma Ray Burst TTE Data }
\label{ex_tte}

\begin{figure}[htb]
\includegraphics[width=5in,height=8in]{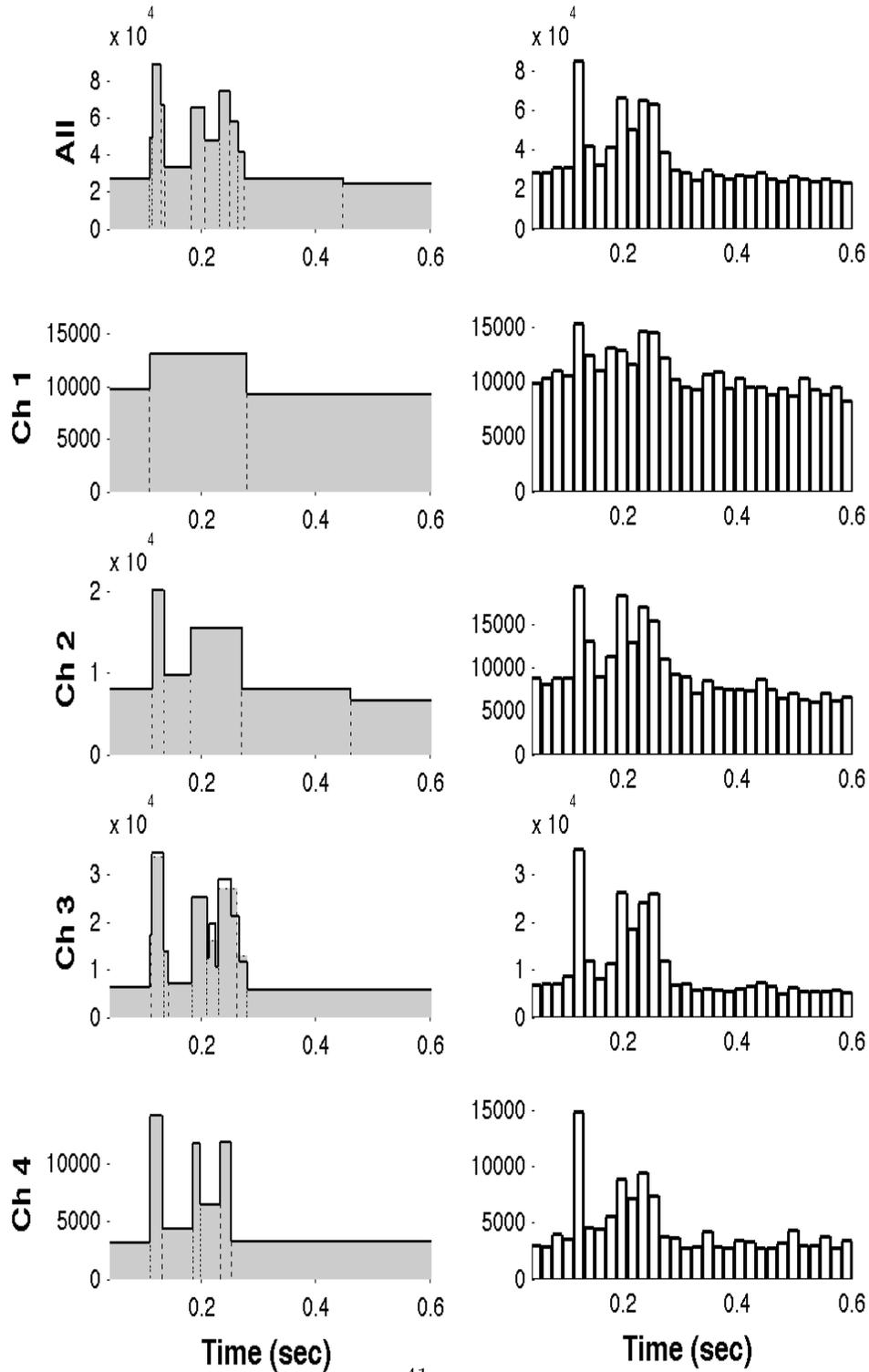} 
\caption{BATSE TTE data for Trigger 0551.
Top panels: all photons.
Other panels: photons in the four BATSE energy channels. 
Left column shows Bayesian Block representations:
default $\mbox{ncp}\_\mbox{prior}$ = solid lines;
iterated $\mbox{ncp}\_\mbox{prior}$ = shaded/dashed lines.
Right column: ordinary evenly spaced binned histograms.
} 
\label{tte_0551}
\end{figure}

\begin{figure}[htb]
\includegraphics[width=5in,height=4.72in]{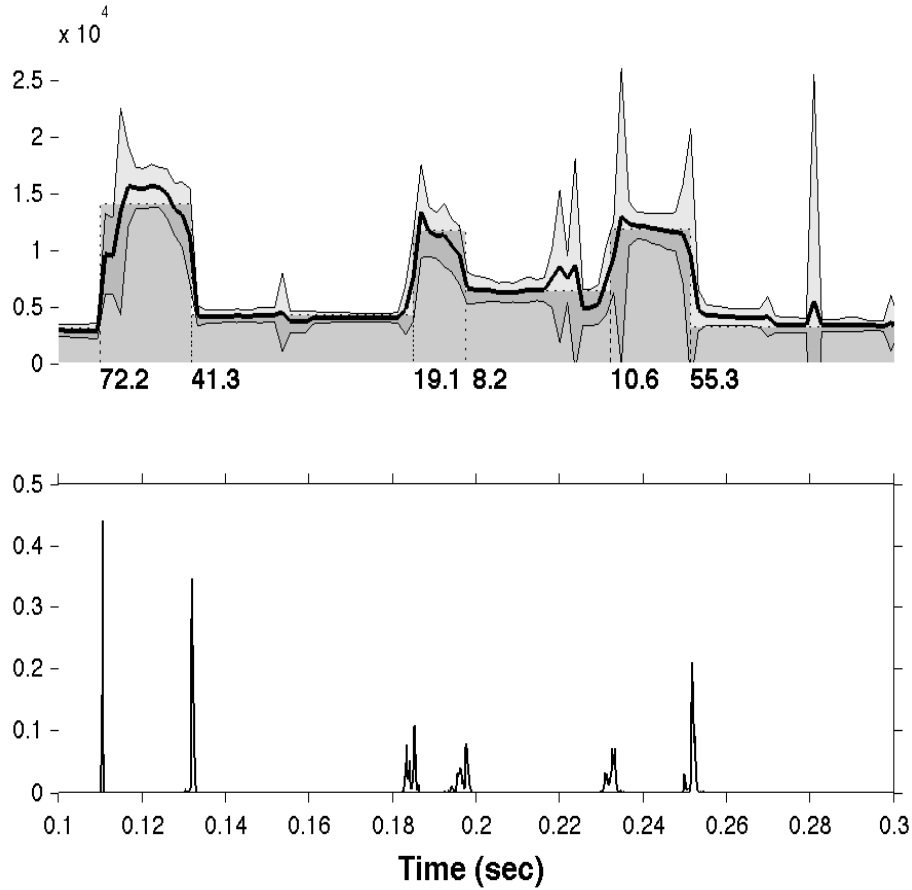} 
\caption{Error analysis for the data in Channel 4 from
Fig. \ref{tte_0551}, zooming in on the time interval
with most of the activity.
Top: Heavy solid line is bootstrap mean (256
realizations), with thin lines giving the $\pm 1 \sigma$
RMS deviations, all superimposed on the BB representation. 
Bottom: approximate posterior distribution functions for the
locations of the change-points, obtained by fixing all of the
others. } 
\label{error_plot}
\end{figure}

Trigger 551 in the BATSE catalog (4B catalog name 910718)
was chosen to exemplify analysis of time-tagged event data
as it has moderate pulse structure.
See \S\ref{determine_prior} for a description of
the data source.
Figure \ref{tte_0551}
shows analysis of all of the event data 
in the top panels, and separated into the
four energy channels in the lower panels.
On the left are optimal block representations
and the right shows the corresponding data 
in 32 evenly spaced bins.

In all five cases 
the optimal block representations 
based on the block fitness function 
for event data in eq. (\ref{fitness_event})
are depicted for two cases, using values
the values of \verb+ncp_prior+:
(1) from eq. (\ref{ncp_fit_event})
with $p_{0} = 0.05$ (solid lines);
and 
(2) 
found with the iterative scheme
described in \S\ref{determine_prior} 
(lightly shaded blocks bounded by dashed lines).
These two results are identical
for all cases except channel 3, 
where the 
iterative scheme's
more conservative control
of false positives 
yields fewer blocks (9 instead of 13).


Note that the ordinary histograms of the
photon times 
in the right-hand panels leave considerable
uncertainty as to what the significant and
true structures are.
In the optimal block representations
two salient conclusions are clear:
(1) there are three pulses, and 
(2) they are most clearly delineated
at higher energies.

\clearpage

This figure depicts the error analysis
procedures
described above in \S\ref{error_analysis}.
\clearpage

\subsection{Multivariate Time Series}
\label{multivariate}

\begin{figure}[htb]
\includegraphics[width=5in,height=5.1in]{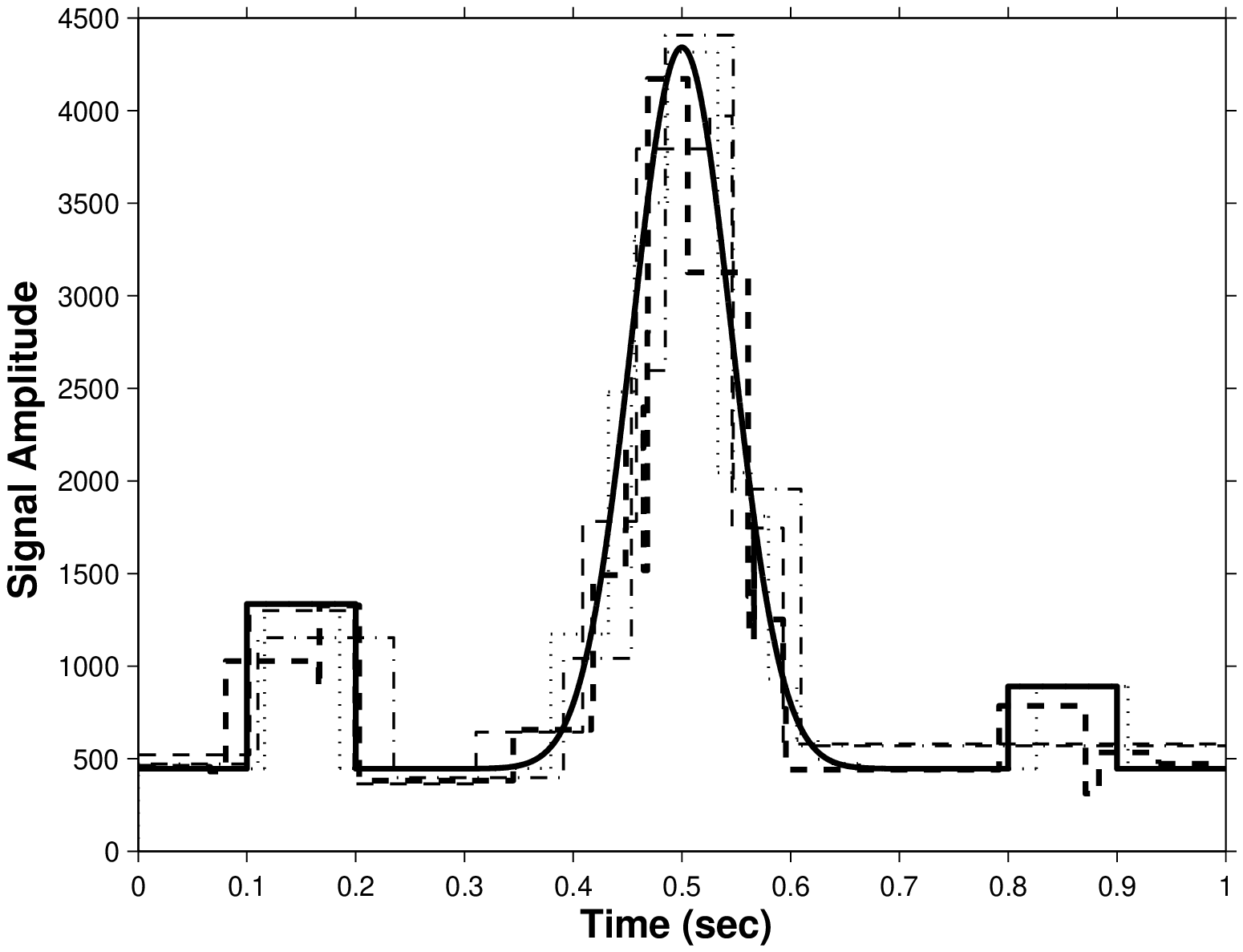} 
\caption{Multivariate analysis of synthetic signal 
consisting of two blocks surrounding 
a Gaussian shape centered
on the interval [0,1] (solid line).
Optimal blocks for three 
independent data series
drawn randomly from the probability 
distribution corresponding to this signal are thin lines:
1024 event times (dash);
4096 events in 32 bins (dot-dash);
and
32 random amplitudes normally distributed 
with mean equal to the signal at
random times uniformly distributed on [0,1]
and constant variance (dots).
The thicker dashed line is the combined
analysis of all three.
} 
\label{mult_fig}
\end{figure}

This example in Fig. \ref{mult_fig} demonstrates 
the multvariate capability of Bayesian Blocks
by analyzing data consisting of three different modes 
sampled randomly from a synthetic signal.
Time-tagged events, binned data, and
normally distributed measurements were
independently drawn from the same signal
and analyzed separately,
yielding the block representations
depicted with thin
lines.

The joint analysis of the 
data combined using the multivariate feature
described above in \S \ref{multi_variate}
is represented as the thick dashed line.
None of these analyses is perfect, of course,
due to the statistical fluctuations in the data.
The combined analysis finds a few
spurious change-points, but overall these
do not represent serious distortions of the true signal.  
The individual analyses are somewhat 
poorer at capturing the true change-points
and only the true change-points.
Hence in this example the combined
analysis makes effective use of disparate
data modes from the same signal.

\subsection{Real Time Analysis: Triggers}
\label{triggers}

Because of its incremental structure 
our algorithm is well suited
for real-time analysis. 
Starting with a small amount of data 
the algorithm typically finds no change-points at first.  
Then by determining the optimal partition 
up to and including the most recently added data cell
the algorithm effectively tests for the
presence of the first change-point. 
The real time mode can be selected 
simply by triggering on the condition  $last(R) > 1$ 
inserted into the code shown in Appendix A , \S\ref{appendix_a},
just before the end of the basic iterative loop on $R$.
For the entry of 1 in each element of array $last$ 
means that the optimal partition consists of 
the whole array encountered so far.
It is thus 
obvious that this first indication of 
change-point cannot yield more than
one change-point.

Thus the algorithm can be set to 
return at the first significant change-point.
Other more complicated 
halting or return conditions 
can be programmed
into the algorithm, 
such as returning after a specified number of change-points
have been found,
or when the location of a change-point 
has not moved for a specified length of time,
etc.  
Essentially any condition on the change-points
or the corresponding blocks can be 
imposed as a halt-and-return condition.

The real time mode 
is mainly of use to 
detect the first sign of a time-dependent
signal rising significantly above a slowly varying background. 
For example, in a photon stream the resulting {\it trigger} 
may indicate
the presence of a new bursting or transient source.

The conventional  way to approach 
problems of this sort 
is to report a
detection if and when the actual event rate, 
averaged over some
interval, exceeds one or more pre-set thresholds.  
See \cite{band}
for an extensive discussion, 
as well as
\cite{fenimore,mclean,schmidt} for other applications 
in high energy
astrophysics. 
One must consider a wide range of configurations:
``BAT uses about 800 different criteria to detect GRBs, each defined
by a large number of commandable parameters.'' \cite{mclean}.  
Both
the size and locations of the intervals over 
which the signal is
averaged affect the result, 
and therefore one must consider many
different values of the corresponding parameters. 
The idea is to
minimize the chances of missing a signal because, for example, its
duration is poorly matched 
to the interval size chosen. 
If the
background is determined dynamically, 
by averaging over an interval
in which it is presumed there is 
no signal, similar considerations
apply to this interval.

Our segmentation algorithm greatly simplifies the
above considerations, since predefined bin sizes and locations are
not needed, and the background is automatically determined in real
time.  
In practice there can be a slight
complication for a continuously 
accumulating data stream, 
since the $N^{2}$ dependence of the
compute time may eventually 
make the computations unfeasible.
A simple countermeasure
is to analyze the data in a sliding window
of moderate size -- large enough to
capture the desired changes but
not so large that the computations take
too long. 
Slow variations in the
background in many cases could 
mandate something like a sliding window anyway.

Because of additional complexities, such as accounting for
background variability and the Pandora's box that spectral
resolution opens \cite{band}, we will defer a serious treatment of triggers to a future publication.

We end with 
a few comments on the
\emph{false alarm} (also called \emph{false positive}) rate
in the context of triggers.
The considerations are very similar to the 
tradeoff 
discussed in the context of 
the choice for the parameter
\verb+ncp_prior+ described 
in \S\ref{ncp_prior},
\S\ref{determine_prior}, 
and \S\ref{fitness_functions} for 
the various data modes.
Even if no signal is present 
a sufficiently large 
(and therefore rare) noise fluctuation can trigger 
any algorithm's detection
criteria. 
Unavoidably all detection procedures embody a trade-off
between sensitivity and rate of false alarms. 
Other things being equal,
making an algorithm more able to 
trigger on weak signals 
renders it more sensitive to noise fluctuations. 
Conversely making an algorithm shun noise
fluctuations renders it insensitive to weak signals. 
In practice  one chooses 
a balance of these competing factors 
based on the relative importance of avoiding
false positives and not missing weak signals. 
Hence there can be no universal prescription.

\subsection{Empty Blocks}
\label{empty_blocks}

Recall that blocks are taken to begin
and end with data cells (\S \ref{change-points}).
This convention means that 
no block can be empty: 
each much contain at least 
its initiating data cell.
Hence in the case of event data, 
blocks cannot represent intervals 
of zero event rate.
This constraint is of no consequence
for the other two data modes.
There is nothing special about zero
(or even negative) signals 
in the case of point measurements.
Zero signal 
would be indicated by 
intervals containing only measured 
values not significantly different from zero.
There is also no issue for binned data
as nothing prevents 
a block from consisting of 
one or more empty data bins.
In many event data applications
zero signal may never occur 
(e.g. if there is a significant background
over the entire observation interval).
But in other cases it may be 
useful to represent such intervals 
in the form of a truly empty block,
with corresponding zero height.

Allowing such null blocks is easily
implemented in a post-processing
step applied to each of the change-points.
The idea is to consider 
reassignment of data cells 
at the start or end of a block 
to the adjoining block 
while leaving the block lengths unchanged.
For a given change-point 
separating a pair of two blocks (``left''
and ``right'')
there are two possibilities:
(a) the datum marking the change-point itself,
currently initiating the right block,
can be moved from the right to the left block;
(b) the datum just prior to the change-point itself,
currently ending the left block, 
can be moved from the left block to the right block.
Straightforward evaluation of the relevant fitness functions
establishes whether one of these
moves increases the fitness of the pair,
and if so which one.
(It is impossible that this calculation 
will favor both moves (a) and (b); 
taken together they yield no
net change and therefore
leave fitness unchanged.)

The suggested procedure is to carry out this
comparison for each change-point
in turn and adjust the populations
of the blocks accordingly.
We have not proved that this
\emph{ad hoc} prescription yields globally 
optimal models with the 
non-emptiness constraint removed,
but it is obvious that the prescription 
can only increase overall model fitness.
It is quite simple computationally and
there is no real downside
to using it routinely, even if 
the moves are almost never triggered.
A code fragment to implement this
procedure is given in Appendix A, \S\ref{appendix_a}.

\subsection{Blocks on the Circle}
\label{circle}

Each of the data spaces discussed so far has been 
a linear interval 
with a well defined beginning and end.
A circle does not have this property.
Our algorithm cannot 
be applied to data defined on a circle,\footnote{Of course 
the case where 
the measured value is confined to a specific subinterval
of the circle is not a problem.}
such as directional measurements, 
because it starts with the first data point 
and iteratively works its way forward along the interval 
to the last point.
Hence the first and last points are 
treated as distant, not as 
the pair of adjacent points that they are.
Any choice of starting point, 
such as the coordinate origin $0$ 
for angles on $[0, 2 \pi ]$, 
disallows the possibility of a block
containing data just before and after it
(on the circle).
In short, 
the iterative (mathematical induction-like) structure
of the algorithm prevents it from being
independent of the 
arbitrary choice of origin,
which on a circle is completely arbitrary.
We have been unable to 
find a solution to 
this problem using 
a direct application of 
dynamical programming.

However there is a method 
that provides exact solutions at the
cost of about one order of magnitude more computation time.
First unfold the data 
with an arbitrary choice for the fiducial origin.
The resulting series starts at this origin, 
continues with the subsequent data points in order, 
and ends at the datum just prior to the fiducial origin. 
Think of cutting a loop of string 
and straightening it out.

The basic algorithm 
is then applied to the data series 
obtained by concatenating
three copies of the unfolded data.
The underlying idea is that the 
central copy 
is insulated from any
effects of the discontinuity 
introduced by the unfolding.
In extensive tests on simulated
data this algorithm performed well.
One check is whether or not 
the two sets of change-points 
adjacent to the two divisions between the 
copies of the data are always 
equivalent (modulo the length of the circle).
These results suggest but do not prove correctness
for all data;
there may be pathological cases 
for which it fails.
Of course this $N^{2}$ computation will 
take $\sim 9$ times as long as it would
if the data were on a simple linear interval.

\begin{figure}[!htb]
\includegraphics[width=5in,height=5in]{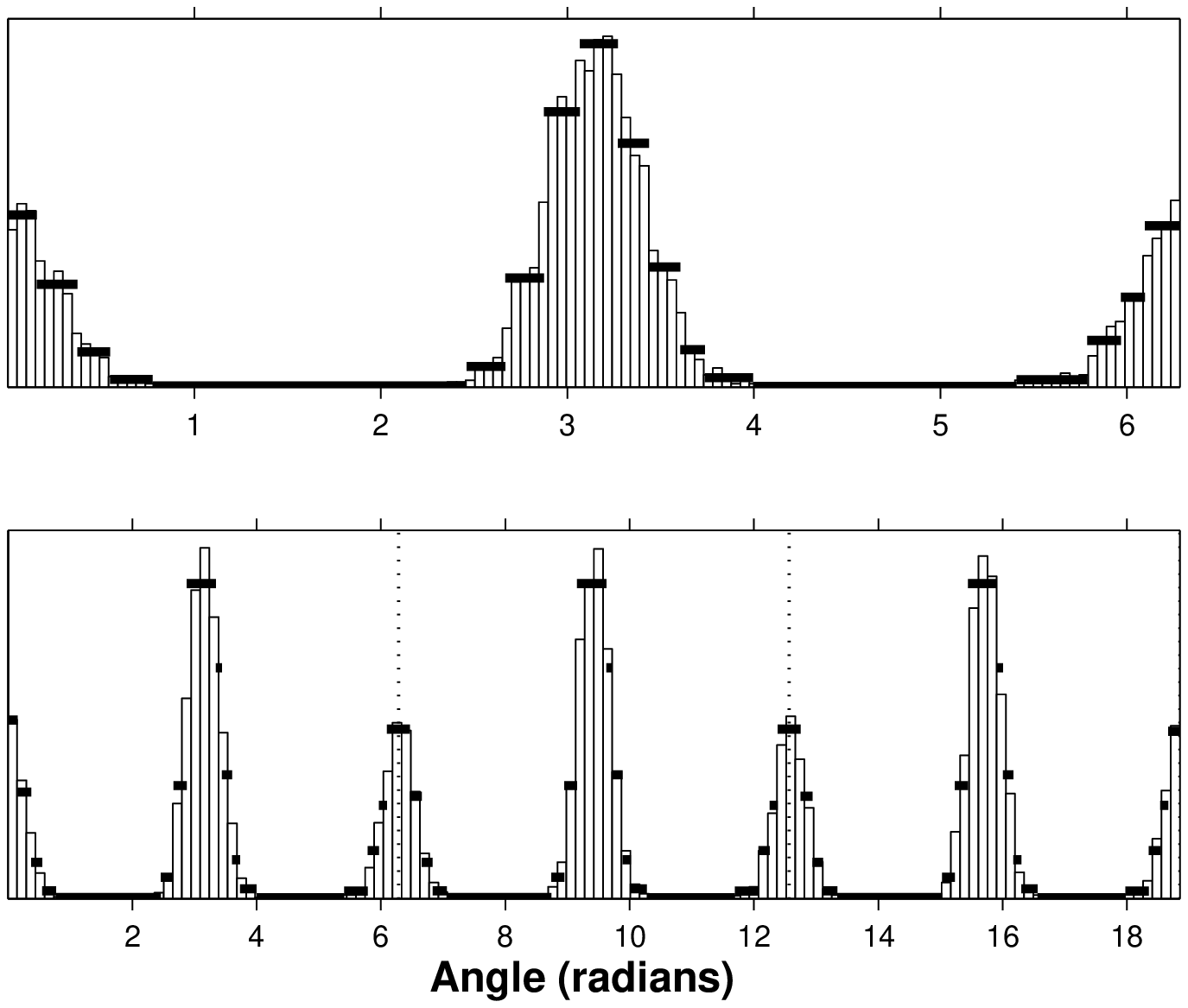}
\caption{Data on the circle: events drawn from two
normal distributions,
centered at $\pi$ and $0$, 
the latter with some points 
wrapping around to values below $2\pi$. 
Optimal blocks are depicted with thick horizontal bars 
superimposed on ordinary histograms.
Top: block representation on the interval [0,2$\pi$].
Bottom: Block representation of 
three concatenated copies 
of the same data on [0,6$\pi$].
Vertical dotted lines
at $2\pi$ and $4\pi$ indicate 
boundaries between the copies.
The blocks in the central copy,
between these lines, 
are not influenced by end effects 
and are the correct optimal representation
of these circular data.  }
\label{circle_fit_1a}
\end{figure}

\begin{figure}[htb]
\includegraphics[width=6in,height=6in]{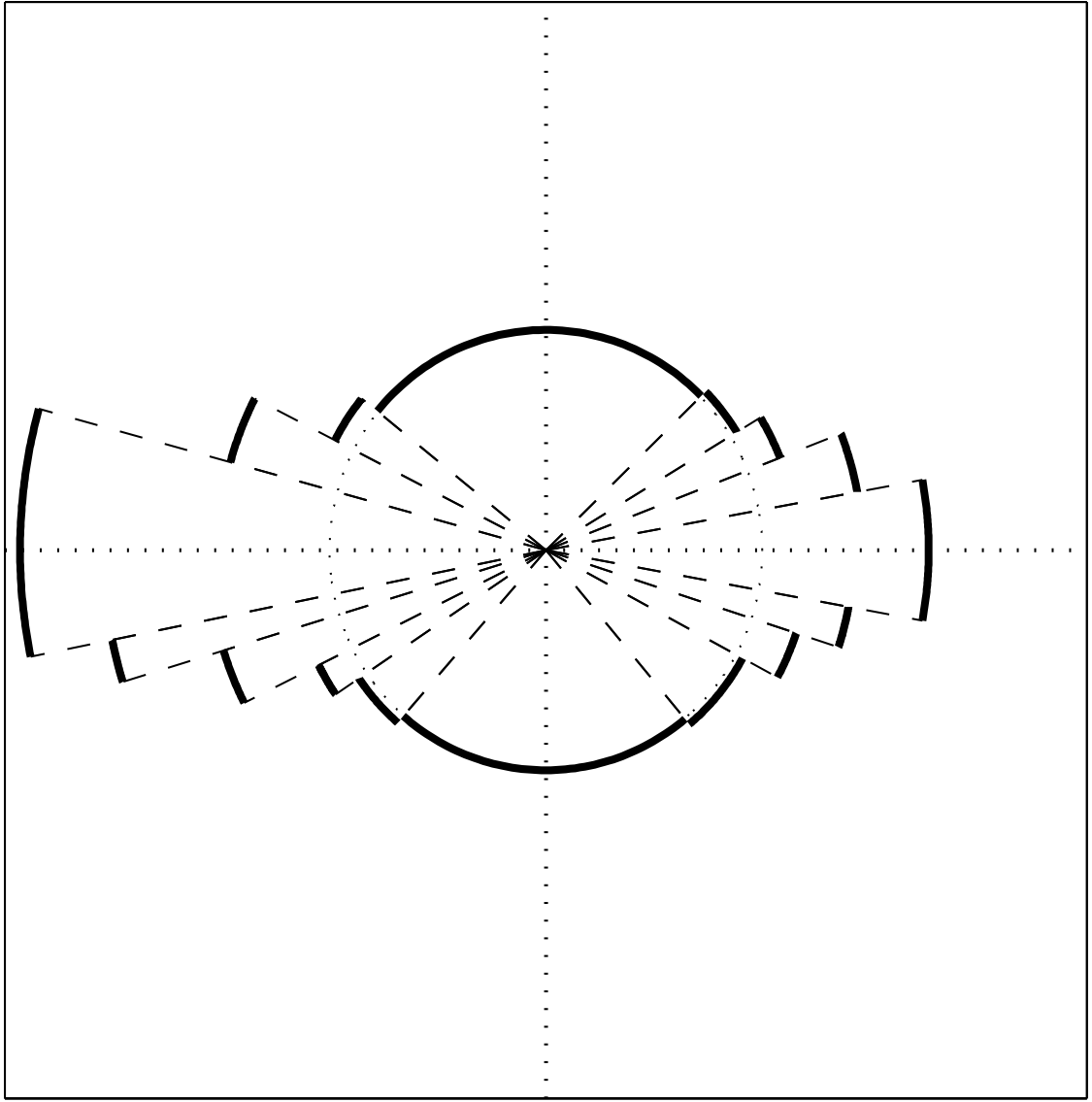}
\caption{Optimal block representation
of the same data as in Figure  \ref{circle_fit_1a} 
(\emph{cf}. the middle third of the bottom panel)
plotted on the circle.
The origin corresponds to the positive x-axis.
and scale of the radius of the circle is arbitrary.}
\label{circle_fit_2}
\end{figure}

Figure \ref{circle_fit_1a} shows simulated data 
representing measurements of an angle on the
interval $[0, 2 \pi]$.
In this case the procedure outlined
above captures the central block (bottom panel) 
straddling the origin that is broken into two
parts if the data series is taken to start at zero
(upper panel).
Note that the two blocks 
just above $0$ and below $2\pi$ in the upper panel,
are rendered as a single block in the central cycle
in the bottom panel.
Figure \ref{circle_fit_2} shows the same 
data shown in Figure \ref{circle_fit_1a} 
plotted explicitly on a circle.
\clearpage

As a footnote, one application
that might not be obvious
is the case of gamma-ray burst light curves
which are short enough that the background
is accurately constant over the duration of the
burst.
If all of the data are rescaled to fit
on a circle, then the pre- and post- 
burst background would automatically
be subsumed into a single block 
(covering intervals 
at the beginning and end of the 
observation period).
This procedure would be 
applicable to  
bursting light curves of any kind
if and only if the background
signal is constant, so that the
event rates before and after
the main burst are the same.
\clearpage

\section{Conclusions and Future Work}
\label{conclusions}

The Bayesian Blocks algorithm 
finds the optimal 
step function model of time series data
by implementing the dynamical
programming algorithm of \cite{jackson}.
It is guaranteed to find the representation 
that maximizes any block-additive fitness function,
in time of order
$N^{2}$,
and replaces the greedy approximate
algorithm in \cite{scargle_v}.
Its real-time mode
triggers on the first 
statistically significant rate change in
a data stream.

This paper addresses 
the following issues in  
the use of the algorithm 
for a variety of data modes:
gaps and exposure variations,
piecewise linear and piecewise exponential models,
the prior distribution for the number of blocks, 
multivariate data, 
the empty block problem (for event data),
data on the circle,
dispersed data, 
and analysis of variance
("error analysis").
The algorithm is shown to closely 
approach the theoretical detection limit 
derived in \cite{adh}.

Work in progress includes 
extensions to generalized data spaces
such as those of higher dimensions (cf. \cite{scargle_4}),
and speeding up the algorithm.

\vskip .5in
Acknowledgements:  This work 
was supported by Joe Bredekamp and the NASA
Applied Information Systems Research Program,
and the CAMCOS program through the Woodward
Fund at San Jose State University.
JDS is grateful for the hospitality of 
the Institute for Pure and Applied Mathematics 
at UCLA, and the Keck Institute for Space Studies
at Cal Tech. We are grateful to 
Glen MacLachlan and Erik Petigura for helpful comments.

\clearpage
\appendix
\section{Reproducible Research: MatLab Code}
\label{appendix_a}
\noindent

This paper implements the spirit of Reproducible Research,
a publication protocol initiated by John Claerbout \cite{claerbout}
and developed by others at Stanford and elsewhere.
The underlying idea is that 
the most effective way of publishing research 
is to include everything necessary to reproduce
all of the results presented in the paper.
In addition to all relevant
mathematical equations and
the reasoning justifying them, 
full implementation of this protocol 
requires that 
the data files and computer programs 
used to prepare all figures and tables
are included.
Cogent arguments for Reproducible Research,
an overview of its development history,
and honest assessment of its successes
and failures, are eloquently 
described in \cite{donoho_rr}.

Following this 
discipline all of the MatLab code and data files 
used in preparing this paper are available 
as auxiliary material.
Included is the file "\verb+read_me.txt+" with details and
a script "\verb+reproduce_figures.m+" 
that erases 
all of the figure files and
regenerates them from scratch.
In some cases the default parameters
implement shorter simulation 
studies than those that were used for the
figures in the paper, 
but one of the features of Reproducible Research 
is that such parameters and other aspects 
of the code can be changed and experimented at will.
Accordingly this collection of scripts 
includes illustrative exemplars of 
the use of and algorithms and 
serves as a tutorial for the methods.

In addition here is
a commented version of 
the key fragment of the MatLab script
(named \verb+find_blocks.m+)
for the basic algorithm described in this paper:

\begin{verbatim}
% For data modes 1 and 2:
% nn_vec is the array of cell populations.
% Preliminary computation:
block_length=tt_stop-[tt_start 0.5*(tt(2:end)+tt(1:end-1))' tt_stop];
...
%-----------------------------------------------------------------
% Start with first data cell; add one cell at each iteration
%-----------------------------------------------------------------
best = [];
last = [];
for R = 1:num_points 
   % Compute fit_vec : fitness of putative last block (end at R)
    if data_mode == 3 % Measurements, normal errors
        sum_x_1 = cumsum( cell_data( R:-1:1, 1 ) )'; %sum(x/sig^2)
        sum_x_0 = cumsum( cell_data( R:-1:1, 2 ) )'; %sum(1/sig^2)
        fit_vec=((sum_x_1(R:-1:1) ) .^ 2 ) ./( 4*sum_x_0(R:-1:1));
    else
        arg_log = block_length(1:R) - block_length(R+1);
        arg_log( find( arg_log <= 0 ) ) = Inf;
        nn_cum_vec = cumsum( nn_vec(R:-1:1) );
        nn_cum_vec = nn_cum_vec(R:-1:1);
        fit_vec = nn_cum_vec .* ( log( nn_cum_vec ) - log( arg_log ) );
    end
    [ best(R), last(R)] = max( [ 0 best ] + fit_vec - ncp_prior );
end
%-----------------------------------------------------------------
% Now find changepoints by iteratively peeling off the last block
%-----------------------------------------------------------------
index = last( num_points );
change_points = [];
while index > 1
    change_points = [ index change_points ];
    index = last( index - 1 );
end
\end{verbatim}

\section{Mathematical Details}
\label{appendix_b}

Partitions of 
arrays of data cells 
are crucial to the block modeling which
our algorithm implements.
This appendix collects a few mathematical
facts about partitions and the nature
of independent events.

\subsection{Definition of Partitions}

A partition of a set is a 
collection of its subsets that add up to the
whole with no overlap.
Formally, a partition is a set of elements, or blocks
$\{B_{k} \}$ satisfying
\begin{equation}\label{completeness}
 I = \bigcup_{k} B_{k}
\end{equation}
and \begin{equation}\label{no_overlap}
B_{j} \bigcap
B_{k} = \emptyset \mbox{ (the empty set) for} \ j \ne k.
\end{equation}
\noindent
Note that these conditions apply to the
partitions of the time series data 
by sets of data cells.
The data cells themselves may or may not 
partition the whole observation interval,
as either the completeness in eq. (\ref{completeness})
or the no-overlap condition in eq. (\ref{no_overlap})
may be violated.

\subsection{Reduction of Infinite Partition Space to a Finite One}

For a continuous independent variable, 
such as time,
the space of all possible partitions is infinitely large.  
We address this difficulty by introducing 
a construct in which 
\interval \ and its partitions are 
represented in terms of a collection of 
$N$ discrete  \emph{data cells}
in one-to-one correspondence with the 
measurements.\footnote{The cells may 
form a partition of \interval,
as for example with event data with no gaps 
(see  \S \ref{event_data}),
but it is not necessary that they do so.} 
The blocks which make up the partitions
are sets of data cells 
contiguous with respect to time-order of the cells.
\emph{I.e.} 
a given block consists of exactly 
all cells with observation times 
within some sub-interval of \interval.

Now consider two sets of partitions of \interval:
(a) all possible partitions 
(b) all possible collections of 
cells into blocks.
Set (a) is infinitely large 
since the  block boundaries 
consist of arbitrary real numbers in \interval,
but set (b) is a finite subset of (a).
Nevertheless, 
under reasonable assumptions 
about the data mode, 
any partition in (a) 
can be obtained from 
some partition in (b) 
by deforming boundaries of 
its blocks without crossing a data
point. 
Because the potential of a block to be an
element of the optimum partition 
(see the discussion of block fitness 
in \S \ref{fitness_functions}) 
is a function of the content of the cells,
such a transformation 
cannot substantially 
change the fitness of the partition.
 
\subsection{The Number of Possible Partitions}

How many different partitions of $N$ cells are
possible? 
Represent a partition by an ordered set of $N$ 
zeros and ones, with one indicating that
the corresponding time is a change-point,
and zero that it is not. 
With two choices at each time, 
the number of combinations is
\begin{equation}
N_{\mbox{partitions}} = 2 ^{ N  } \ .
\end{equation}
\noindent
Except for very short time series this number is
too large for an exhaustive search, but
our algorithm nevertheless finds the optimum
over this space
in a time that scales as only $N^{2}$.

\subsection{A Result for Subpartitions}
\label{subpartitions}
We here define \emph{subpartitions} and prove
an elementary corollary that is key to the algorithm.
\vskip 0.2in
\begin{center}
\noindent \framebox[6.in][l]{\parbox{6.in}
{\center{{\bf Definition:} a {\it subpartition} of a given partition 
$\mbox{\MyScript P}(I)$ \\
is 
a subset of the blocks of $\mbox{\MyScript P}(I)$. \vskip .2in} } }
\end{center}
\vskip 0.25in \noindent 
It is obvious that a subpartition is a partition of that subset
of \interval \ consisting of those blocks.
Although not a necessary
condition for the result to be true, in all cases of interest here
the blocks in the subpartition are contiguous, and thus form a
partition of a subinterval of \interval. 
It follows that: \vskip 0.2in
\begin{center}
\noindent \framebox[6.in][l]{\parbox{6.in}
{ \center{{\bf Theorem:} A subpartition $\mbox{\MyScript P}'$ of 
an optimal partition $\mbox{\MyScript P}(I)$} \\
is an optimal partition of the
subset $I'$ that it covers.
\vskip .2in }}
\end{center}
\vskip 0.25in \noindent 
For if there
were a partition of $I'$, different from and fitter than
$\mbox{\MyScript P}'$, then combining it with the blocks of
$\mbox{\MyScript P}$ not in $\mbox{\MyScript P}'$ would, by the
block additivity condition, yield a partition of \interval  fitter than
$\mbox{\MyScript P}$, 
contrary to the optimality of $\mbox{\MyScript P}$. 

We will make use of 
the following corollary:
\vskip .2in
\noindent \framebox[6.in][l]{\parbox{6.in}
{\center {\bf Corollary}: removing the last block 
of an optimal partition \\
leaves an optimal partition. \vskip 0.2in }}
\vskip 0.25in \noindent 

\subsection{Essential Nature of the ``Poisson'' Process}
\label{poisson_nature}

The term \emph{Poisson process} refers to events 
occurring randomly in time 
and \emph{independently of each other}.
That is, the times of the events,
\begin{equation}
t_{n}, n = 1, 2, \dots , N \ ,
\end{equation}
\noindent
are independently 
drawn from a given probability distribution.
Think of the events as darts thrown 
randomly at the interval.
If the distribution is flat 
(\emph{i.e.} the same all over the interval of interest)
we have a \emph{constant rate Poisson process}.
In this special case 
a point is just as likely to occur anywhere in
the interval as it is anywhere else;
but this need not be so. 
What must be so in general -- the essential nature 
of the Poisson process from a physical 
point of view -- is the above-mentioned
independence: each dart is
not at all influenced by the others.
Throwing darts that have feathers or magnets,
although random, is not a Poisson process
if these accoutrements cause the darts to 
repel or attract each other.

This key property of independence determines 
all of the other features of the process.
Most important are a set of remarkable
properties of interval distributions
(see \emph{e.g.} \cite{papoulis}).  
The time interval 
between a given point $t_{0}$
and the time $t$ of the next event
 is exponentially distributed
\begin{equation}
P( \tau ) d\tau = \lambda e^{ - \lambda \tau } d\tau \ ,
\label{poisson_intervals}
\end{equation}
\noindent
where  $\tau = t - t_{0}$.
The remarkable aspect is that 
it does not matter how $t_{0}$ is chosen;
in particular the distribution is the
same whether or not an event occurs at
$t_{0}$.  
This fact makes the implementation 
of event-by-event exposure straightforward
(\S\ref{exposure_variations}).

Note that we have not mentioned 
the Poisson distribution itself.
The number of events in a fixed 
interval does obey the Poisson distribution,
but this result is subsidiary to,
and follows from,
event independence.
In this sense
a better name than \emph{Poisson process} is 
\emph{independent event process}.

In representing intensities of such processes,
one scheme is to represent
each event as  a delta-function in time. 
But a more convenient way 
to extract rate information incorporates 
the time intervals\footnote{A method
for analyzing event data based solely
on inter-event time intervals has
been developed in (\cite{prahl}).}
between photons.
Specifically, for each photon
consider the interval starting half way
back to the previous photon and ending 
half way forward to the subsequent photon.
This interval, namely 
\begin{equation}
[ {t_{n} - t_{n-1} \over 2}, {t_{n+1} - t_{n} \over 2} ]\ , 
\label{interval_2}
\end{equation}
\noindent
is the set of times closer to $t_{n}$ than
to any other time,\footnote{These intervals form the
{\emph Voronoi tessellation} of the
total observation interval. See (\cite{spatial_tessellations})
for a full discussion of this construct,
highly useful in spatial domains of 2, 3, or
higher dimension; see also (\cite{scargle_2,scargle_4}).}
and has length equal to the average of the
two intervals connected by photon $n$, namely 
\begin{equation}
 \Delta t_{n} = { t_{n+1} - t_{n-1} \over 2} \ .
\end{equation}
Then the reciprocal 
\begin{equation}\label{reciprocal}
x_{n} \equiv { 1 \over \Delta t_{n} }
\end{equation}
is taken as an estimate
of the signal amplitude corresponding 
to observation $n$.
When the photon rate 
is large, the corresponding intervals are small.
\begin{figure}
\includegraphics[width=4in,height=6in]{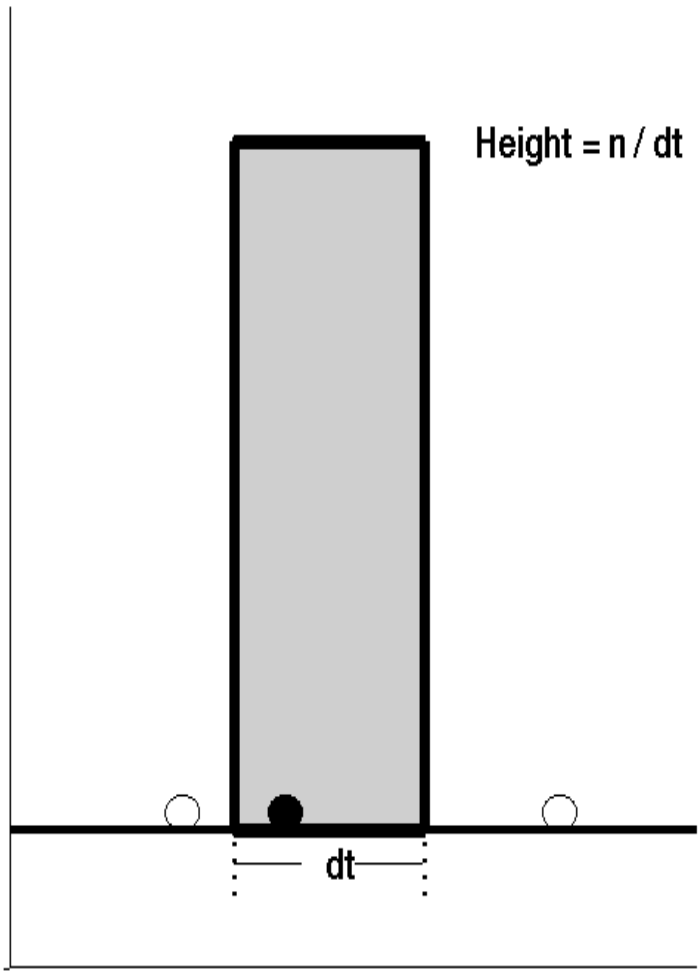}
\caption{Voronoi cell of a photon.
Three successive photon detection times are 
circles on the time axis.  
The vertical
dotted lines underneath delineate the 
time extent ($dt$) of the cell 
and the height of the rectangle --
${n / dt}$,
where $n$ is the number of photons at exactly 
the same time (almost always 1) -- 
is the local estimate of the
signal amplitude.  
If the exposure at this
time is less than unity, 
the width of the
rectangle shrinks in proportion,
the area of the rectangle is preserved, 
so the height increases 
in inverse proportion
yielding a larger estimate of the true event rate.}
\label{voronoi_cell}
\end{figure}
demonstrates the data cell concept,
including the simple modifications to
account for variable exposure 
and for weighting
by photon energy.

\cite{prahl} has derived a statistic for event clustering
in Poisson process data that tests departures from the known
interval distribution by evaluating the
likelihood over a restricted interval range. Prahl's statistic is
\begin{equation}
    M_{N} = {1 \over N} \sum_{\Delta T_{i} < C^{*} }
    (1 - {\Delta T_{i} \over C^{*} }
    ) \ ,
\label{prahl_1a}
\end{equation}
\noindent where $\Delta T_{i}$ is the interval between events $i$
and $i+1$, and
\begin{equation}\label{prahl_2}
C^{*} \equiv  {1 \over N} \sum \Delta T_{i}
\end{equation}
is the empirical mean interval.  In other settings, the fact that
this statistic is a global measure of departure of the distribution
(used here only locally, over one block) may be useful in the
detection of periodic, and other global, signals in event data.

\section{Other Block Fitness Functions}
\label{appendix_c}

This appendix describes fitness function for
a variety of data modes.

\subsection{Event Data: Alternate Derivation}
\label{fitness_events_alt}

The Cash statistics used to derive 
the fitness function in Eq. (\ref{fitness_event})
is based on representation of event
times as real numbers.
Of course time is not measured with 
infinite precision, so it is interesting 
to note that a more realistic treatment
yields the same formula.

Typically the data systems' 
finest time resolution is represented
as an elementary quantum of time,
which will be called a
\emph{tick} since it is usually set by a
computer clock.
Measured values are expressed 
as integer multiples of it
(\emph{cf.}  \S 2.2.1 of \cite{scargle_v}).
We assume that $n_{m}$, 
the number of events (\emph{e.g.} photons) 
detected in tick $m$ 
obeys a Poisson distribution:
\begin{equation}\label{like_2}
    L_{m} = { (\lambda dt) ^{n_{m}} \ e^{ -\lambda dt }  \over n_{m}! } 
    = { \Lambda ^{n_{m}} \ e^{ -\Lambda }  \over n_{m}! } \ ,
\end{equation}
\noindent 
where 
$dt$ is the length of the tick.
The event rates
$\lambda$ and $\Lambda$ are 
counts per second
and per tick, respectively.
Time here is given 
in units such as seconds,
but a representation in terms of 
(dimensionless) integer multiples of $dt$
is sometimes more convenient.

Due to event independence
the block likelihood is the product of these individual factors
over all ticks in the block. 
Assuming all ticks have the same length $dt$ this is:
\begin{equation}\label{block_like_2a}
    L^{(k)} = \prod_{m=1}^{M^{(k)}}
    { (\lambda dt )^{n_{m} } e^{- \lambda dt } \over n_{m}! } \ ,
\end{equation}
\noindent where $M^{(k)}$ is the number of ticks in block $k$.
Note that non-events 
are included via the factor $e^{- \lambda dt }$ for each 
tick with $n_{m} = 0$.
When this expression is used to 
compute the likelihood for the whole interval 
(\emph{i.e}. product of the block likelihoods over 
all blocks of the model) 
the denominator contributes the factor
\begin{equation}\label{factorial_factor}
{1 \over
     \prod_{k} \prod_{m}^{ M^{(k)} }
    {n_{m}! } } 
    =
     {1 \over
     \prod_{m}
    {n_{m}! }} 
    \ ,
\end{equation}
\noindent
where on the right-hand side 
the product is over all the ticks in the whole interval.
For low event rates where $n_{m}$
never exceeds $1$, 
this quantity is unity.
No matter what
it is a constant, fixed once and for all 
given the data;
in model comparison contexts 
it is independent of model parameters
and hence irrelevant.
Dropping it, noting that 
$\prod_{m=1}^{ M^{(k)}}  e^{- \lambda dt }$ is
just
$e^{-\lambda M^{(k)} dt } = e^{-\lambda M^{(k)} }$ 
Collecting together all factors for ticks with the same
number of events 
eq. (\ref{block_like_2a}) simplifies to 
\begin{equation}\label{block_like_2d}
    L^{(k)} =  e^{-\lambda M^{(k)} } \prod_{n=0}^{\infty}
    ( \lambda dt ) ^{n H^{(k)}(n)}
    \ ,
\end{equation}
\noindent where $H^{(k)}(n)$ is the number of ticks in the block with $n$
events. Noting that
\begin{equation}
\sum_{n=0}^{\infty} n H^{(k)}(n) = N^{(k)} \ , 
\label{block_count}
\end{equation}
where $N^{(k)}$ is the total number of events in 
block $k$,
we have simply
\begin{equation}
\label{block_like_2e}
    L^{(k)} = (\lambda dt )^{N^{(k)} } e^{- \lambda M^{(k)} } \ .
\end{equation}

In order for the model to depend on 
only the parameters defining the block edges, 
we need to eliminate
$\lambda$ from eq. (\ref{block_like_2e}).
One way to do this is to find the maximum of
this likelihood as a function of $\lambda$, 
which is easily seen to be at 
$\lambda = {N^{(k)} \over M^{(k)}}$,
yielding 
\begin{equation}\label{maxed_like_2}
    L_{max}^{(k)}  
    =
( {N^{(k)} dt \over M^{(k)}} )^{N^{(k)} } e^{- N^{(k)} }
\end{equation}
\noindent 
The exponential contributes the 
overall constant factor 
$e^{- \sum_{k} N^{(k)} } =  e^{- N}$
to the full model.
Moving 
this ultimately irrelevant factor 
to the left-hand side,
noting that $M^{(k)} = {T^{(k)}  \over dt}$, 
and taking the log, we have
for the maximum-likelihood
block fitness function
\begin{equation} 
\label{ml_fitness_alt}
\mbox{log} \ L^{ (k) }_{max} + N^{(k)}  = 
 N^{(k)} ( \  \mbox{log} N^{(k)} - \mbox{log}  M^{(k)} ) 
 \ 
 \ .
 \end{equation}
 \noindent
equivalent to Eq. (\ref{fitness_event}).

An alternative way to eliminate $\lambda$
is to marginalize it as in the Bayesian
formalism.   
That is, one specifies a prior probability
distribution for the parameter and
integrates the likelihood in Eq. 
(\ref{block_like_2e}) times this prior.
Since the current context is generic,
not devoted to a specific application,
we seek a distribution that expresses
no particular prior knowledge for the
value of $\lambda$.  It is well known
that there are several practical and
philosophical issues connected with
such so-called \emph{non-informative} priors.
Here we adopt this simple flat, normalized prior:
\begin{equation}\label{flat_prior_normalizable}
 P(\lambda) = {\Huge \{ }
  \begin{tabular}{@{} cc @{}}
    $P^{(\Delta)}$  & $\lambda_{1} \le \lambda  \le \lambda_{2}$ \\ 
    0 & \mbox{otherwise} \\ 
  \end{tabular} \ \ ,
\end{equation}
\noindent
where the normalization condition yields
\begin{equation}
P^{(\Delta)} = { 1 \over \lambda_{2} - \lambda_{1} } 
 = {1 \over \Delta \lambda } \ .
\end{equation}
\noindent
Thus eq. (\ref{block_like_2e}),
with $\lambda$ marginalized,
is the posterior probability 
\begin{eqnarray}\label{block_like_2f}
  P_{\mbox{\small marg}}^{(k)} &= P^{(\Delta)}
    \int_{\lambda_{1}}^{\lambda_{2}}
    ( \lambda dt)^{N^{(k)} } e^{- \lambda T^{(k)} }
    d\lambda \\
     &= { P^{(\Delta)} \over T^{(k)}}
      ({dt   \over   T^{(k)} }) ^{N^{(k)}}
    \int_{z_{1}}^{z_{2}}
     z^{N^{(k)} } e^{-z } dz
\end{eqnarray}
\noindent 
where $z_{1,2} = T^{(k)}\lambda_{1,2}$.  
In terms of 
the \emph{incomplete gamma function}
\begin{equation}\label{igf}
\gamma( a, x ) \equiv
 \int_{0}^{x} z^{a-1} e^{-z} dz \ ,
\end{equation}
\noindent
we have, utilizing $M^{k} = {  T^{(k)} \over dt  }$, 
\begin{equation}\label{mode_2_post}\fbox{$ \ \ \
 \mbox{log}  P_{\mbox{\small marg}} ^{(k)} = 
 \mbox{log} { P^{(\Delta)} \over T^{(k)}}
     - N^{(k)} \mbox{log} M^{(k)} 
      + \mbox{log} [\ \gamma( N^{(k)} + 1, z_{2} ) -
     \gamma( N^{(k)} + 1, z_{1} )\ ]  \ \ \ $}  \ .
\end{equation}
\noindent The infinite range 
  $z_{1} = 0, z_{2} = \infty$,
gives
\begin{equation}\label{mode_2_post_inf}\fbox{$ \ \ \
  \mbox{log} P_{\mbox{\small marg}(\infty)}^{(k)} = \mbox{log} { P^{(\Delta)} \over T^{(k)}}
    + \mbox{log} \Gamma( N^{(k)} + 1 )  
  - N^{(k)} \mbox{log} M^{(k)} 
  \ \ \ 
  $} 
  \   ,
\end{equation}
\noindent
This prior 
is unnormalized (and therefore
sometimes regarded as improper).
Technically $P^{(\Delta)}$ approaches
zero
as $z_{2} \rightarrow \infty$,
but is retained here
in order to formally retain
the scale invariance 
to be discussed at the end
of this section.

\vskip 0.25in
Another commonly used prior is the so-called conjugate Poisson
distribution
\begin{equation}
\label{conj_prior}
P( \lambda ) = C \ \lambda^{\alpha - 1} e^{-\beta \lambda } \ .
\end{equation}
\noindent As noted by \cite{gelman} this ``prior density is, in some
sense, equivalent to a total count of $\alpha$-1 in $\beta$ prior
observations,'' a relation that might be useful in some
circumstances.  
The normalization constant 
$C = {\beta^{\alpha},\over \Gamma(\alpha)}$,
and with this prior the
marginalized posterior probability distribution is
\begin{equation}
P_{\mbox{\small cp}} = 
C \int_{0}^{\infty} \lambda^{N^{(k)} + \alpha - 1} e ^{ - \lambda
( M^{(k)} + \beta ) } d \lambda \ ,
\end{equation}
\noindent yielding 
\begin{equation}\label{binned_posterior_3}\fbox{$
\mbox{log} \ P_{\mbox{\small cp}}  - \mbox{log} \ C =
 \mbox{log} \ {\Gamma(N^{(k)} + \alpha) 
 - ( N^{(k)} + \alpha }) \  \mbox{log} ( M^{(k)} + \beta ) $} \ .
\end{equation}
\noindent Note that for $\alpha = 1, \beta = 1$
this prior and posterior reduce to those in Eqs.
(28) and (29) of \cite{scargle_v}.

Equations (\ref{fitness_event}), (\ref{mode_2_post}),
(\ref{mode_2_post_inf}) and (\ref{binned_posterior_3})
are all invariant under a
change in the units of time.
The case of
eq. (\ref{mode_2_post_inf})
is slightly dodgy,
as mentioned above,
but otherwise is a direct result of
expressing 
$N^{(k)}$ and $M^{(k)}$
as dimensionless counts,
of events and time-ticks, respectively.
(Further, in the
case of eq. (\ref{mode_2_post}),
$z_{1}$ and $z_{2}$ are dimensionless.)
As mentioned above, the simplicity of
eq. (\ref{fitness_event}) recommends it
in general, but specific prior information
(e.g. as represented by eq. \ref{conj_prior})
may suggest use of one of the other forms.


\subsection{0-1 Event Data: Duplicate Time Tags Forbidden}
\label{event_mode_2}

In Mode 2 duplicate time tags are not allowed, the number of
events detected at a given tick is $0$ or $1$, and the corresponding
tick likelihood is:
\begin{eqnarray}
  L_{m} & = e^{ -\lambda dt } = 1 - p  \ \ \ \ \ & n_{m} = 0 \ \ \  \\
                   & = 1 - e^{-\lambda dt } = p  \ \ \ \ \ \ \ & n_{m} = 1 \  \ \  \label{like_1}
\end{eqnarray}
\noindent where $\lambda$ is the model event rate,
in events per unit time. 
From the Poisson
distribution $p = 1 - e^{-\lambda dt }$ is 
the probability of an event,
and
$1 - p = e^{-\lambda dt }$ that of no event. 
Note that $p$
or $\lambda$ interchangeably specify the event rate. 
Since
independent probabilities multiply, the block likelihood is the
product of the tick likelihoods:
\begin{equation}\label{block_like_1a}
    L^{(k)} = \prod_{m=1}^{M^{(k)}} L_{m} =
    p^{N^{(k)}} (1-p) ^{ M^{(k)} - N^{(k)} } \ , 
\end{equation}
\noindent where $M^{(k)}$ is the number of ticks in block $k$ and
$N^{(k)}$ is the number of events in the block.

There are again two ways to proceed.  The maximum of this likelihood
occurs at $p = { N^{(k)} \over M^{(k)} }$ and is
\begin{equation}\label{maxed_like}
    L_{max}^{(k)} = ( { N^{(k)} \over M^{(k)}} )^{N^{(k)}}
     (1-{N^{(k)} \over M^{(k)} }) ^{ M^{(k)} - N^{(k)} }
\end{equation}
\noindent  Using the logarithm of the maximum likelihood,
\begin{equation}\fbox{$
 \mbox{log}  L_{max} ^{(k)}  = {N^{(k)}} 
 \mbox{log} ( { N^{(k)} \over M^{(k)}} ) +
     ( M^{(k)} - N^{(k)} ) \mbox{log}  (1-{N^{(k)} \over M^{(k)} }) $}
\label{mode_2_1}
\end{equation}
\noindent
yields the fitness function, additive over blocks.

As in the previous sub-section,
an alternative is to marginalize 
$\lambda$:
\begin{equation}\label{mode_1_margin}
    P^{(k)} = \int L^{(k)} P( \lambda ) d\lambda \ ,
\end{equation}
\noindent where $P( \lambda )$ is the prior probability distribution
for the rate parameter. 
With the flat prior in eq. 
(\ref{flat_prior_normalizable})\footnote{In
\cite{scargle_v} we used $p$ as the independent variable, and chose
a prior flat (constant) as a function of $p$.  Here, we use a prior
flat as a function of the rate parameter.}
the posterior, marginalized over $\lambda$ is
\begin{equation}\label{mode_1_marg_int}
    P_{\mbox{\small marg}}^{(k)} =  P^{(\Delta)}
    \int_{\lambda_{1}}^{\lambda_{2}}
    (1 - e^{-\lambda dt })^{N^{k}} 
    (e^{-\lambda dt }) ^{ M^{(k)} - N^{k} }
    d\lambda \ .
\end{equation}
Changing variables to $p = 1 - e^{-\lambda  dt}$, 
with $dp = dt \ e^{-\lambda dt } d\lambda$, 
this integral becomes
\begin{equation}\label{mode_1_marg_int_1}
 P_{\mbox{\small marg}}^{(k)}  = { P^{(\Delta)} \over dt} 
    \int_{p_{1}}^{p_{2}}
    p^{N^{(k)}} (1-p) ^{ M^{(k)} - N^{(k)}- 1 }
    dp \ ,
\end{equation}
with $p_{1,2} = 1 - e^{-\lambda_{1,2}  dt}$, 
and 
expressible in terms of the \emph{incomplete
beta function}
\begin{equation}\label{incomp_beta}
   B(z; a, b) = \int_{0}^{z} u^{a-1} ( 1 - u ) ^{b-1} du
\end{equation}
as follows:
\begin{equation}\label{mode_2_2}\fbox{$
\mbox{log} P_{\mbox{\small marg}}^{(k)}  -  
\mbox{log} { P^{(\Delta)} \over dt}
= \newline
\mbox{log}
    [ B(p_{2}; N^{(k)}+1,M^{(k)}-N^{(k)})-
     B(p_{1}; N^{(k)}+1,M^{(k)}-N^{(k)}) ]\ .$}
\end{equation}
The case $p_{1} = 0, p_{2} = 1$ 
yields the ordinary \emph{beta
function}:
\begin{equation}\label{mode_2_3}\fbox{$
\mbox{log} 
P_{0 \rightarrow 1}^{(k)}  - \mbox{log} { P^{(\Delta)} \over dt}
= \mbox{log} 
B( N^{(k)}+1, M^{(k)} - N^{(k)} )\ , $}
\end{equation}
\noindent differing from Eq. (21) of \cite{scargle_v} by one in the
second argument, due to the difference between a prior flat in $p$
and one flat in $\lambda$.  All of the equations (\ref{mode_2_1}),
(\ref{mode_2_2}), and (\ref{mode_2_3}), 
can be used as fitness functions in the global optimization
algorithm and, 
as with Mode 1, 
are invariant to a change in the units of time.

\vskip 0.25in
\hrule
\vskip 0.25in

A brief aside:
one might be tempted to use 
intervals between successive events 
instead of the actual times,
since in some sense they express
rate information more directly.  
However, as we now prove,
the likelihood based on intervals is essentially
equivalent to that in eq. (\ref{block_like_2e}).
It is a classic result \cite{papoulis}
that intervals between (time-ordered) consecutive 
independent events 
(occurring with a probability uniform in
time, with a constant rate $\lambda$) 
are exponentially distributed:
\begin{equation}\label{intervals}
    P( dt ) dt = \lambda e^{- \lambda dt } U( dt ) dt,
\end{equation}
\noindent where $U(x)$ is the unit step function:
\begin{eqnarray*}
  U(x)&=&1 \ \ \ \ \ x \ge 0 \\
  \ &=&0 \ \ \ \ \ x < 0 \ .
\end{eqnarray*}
\noindent Pretend that the data consists of the inter-event
intervals, 
and that one does not even know the absolute times. The
likelihood of our constant-rate Poisson model for interval $dt_{n}
\ge 0$ is
\begin{equation}\label{intervals_1}
L_{n} = \lambda e^{- \lambda \ dt_{n} },
\end{equation}
\noindent 
so the block likelihood is
\begin{equation}
\label{intervals_2}
L^{(k)} = \prod_{n=1}^{N^{(k)}} \lambda \ e^{- \lambda \ dt_{n} } =
\lambda^{N^{(k)}} e^{- \lambda M^{(k)} },
\end{equation}
the same as in eq. (\ref{block_like_2e}),
except that here $N^{(k)}$ is the number of 
inter-event
intervals, one less than the number of events.

\cite{prahl} derived a statistic 
for event clustering,
by testing for significant departures from the known
interval distribution, by evaluating the
likelihood over a restricted interval range. This statistic is
\begin{equation}
    M_{N} = {1 \over N} \sum_{\Delta T_{i} < C^{*} }
    (1 - {\Delta T_{i} \over C^{*} }
    ) \ ,
\label{prahl_3}
\end{equation}
\noindent where $\Delta T_{i}$ is the interval between events \interval \
and $i+1$, 
$N$ is the number of terms in the sum,
and
\begin{equation}
C^{*} \equiv  {1 \over N} \sum \Delta T_{i}
\label{prahl_1}
\end{equation}
is the empirical mean 
of the relevant intervals.  
In some settings, the fact that
this statistic is a global measure 
(as opposed to the local -- over one block at a time -- ones used here) may be useful in the
detection of global signals, such as periodicities,
in event data.

\vskip 0.25in
\hrule
\vskip 0.25in

\subsection{Time-to-Spill Data}

As discussed in \S 2.2.3 of \cite{scargle_v}, 
reduction of the necessary telemetry rate is sometimes accomplished by recording only
the time of detection of every Sth photon, e.g. with S=64 for the
BATSE time-to-spill mode.  
This data mode has the attractive feature
that its time resolution is greater when the source is brighter (and
possibly more active, so that more time resolution is useful).  
With slightly revised notation the
likelihood in Eq. (32) of \cite{scargle_v} simplifies to
\begin{equation}\label{tts_likelihood}
    L_{TTS}^{(k)} = \lambda ^{SN_{\mbox{\small spill}}^{(k)}} e^{- \lambda M^{(k)} }
\end{equation}
\noindent where $N_{\mbox{\small spill}}^{(k)}$ is the number of spill events in the
block, and $M^{(k)}$ is as usual the length of the block
in ticks.  With $N
= N_{\mbox{\small spill}}^{(k)} S$ this is identical to the Poisson likelihood in
Eq.(\ref{block_like_2a}), and in particular the maximum likelihood is at
$\lambda = {N_{\mbox{\small spill}}^{(k)}S \over M^{(k)}}$ and the corresponding fitness
function is
\begin{equation}\label{tts_ml}
log L_{max, TTS}^{(k)} - \mbox{log} N 
= SN_{\mbox{\small spill}}^{(k)} \ 
[ \ \mbox{log} (N_{\mbox{\small spill}}^{(k)}S )  - \mbox{log} M^{(k)} \ ]
\end{equation}
just as in Eq. (\ref{fitness_event}) with $N^{(k)} = SN_{\mbox{\small spill}}^{(k)}$, and with
the same property that the unit in which block lengths are expressed
is irrelevant.

\subsection{Point measurements: 
Alternative Form}
\label{point_1}
An alternative form can be derived
by inserting (\ref{lam_max_wx}) 
instead of (\ref{lambda_max_2})
into the log of Eq. (\ref{block_like})
as in \S \ref{point_measurements}.
The result is:
\begin{equation}
\mbox{log} L^{(k)}_{\mbox{\small max}} =
- {1 \over 2} \sum_{n} ({ x_{n} - \sum_{n'} w_{n'} x_{n'} 
\over \sigma_{n} })^2 
\label{log_like}
\end{equation}
\noindent 
Expanding the square gives
\begin{equation}
\mbox{log} L^{(k)}_{\mbox{\small max}} = 
- {1 \over 2} [ \ 
 \sum_{n} ({ x_{n}  \over \sigma_{n} })^2 
 -2 
 \sum_{n} ( { x_{n} \over \sigma_{n}^{2} }) ( \sum_{n'} w_{n'} x_{n'} )
 + ( \sum_{n'} w_{n'} x_{n'} )^{2} 
 \sum_{n} {1 \over \sigma_{n}^{2} }  \  ]
\label{log_like_a}
\end{equation}
\noindent 

\begin{equation}
 = 
- {1 \over 2} \sum_{n'}( {1 \over \sigma_{n'}^{2} } )  [ \ 
 \sum_{n} w_{n} x_{n}^{2}
 -2 
 (\sum_{n} w_{n} x_{n}) ( \sum_{n'} w_{n'} x_{n'} )
 + ( \sum_{n'} w_{n'} x_{n'} )^{2} 
 \sum_{n} w_{n}  \  ]
\label{log_like_1a}
\end{equation}

\noindent 
\begin{equation}
 =  
- {1 \over 2} \sum_{n'}( {1 \over \sigma_{n'}^{2} } )  [ \ 
 \sum_{n} w_{n} x_{n}^{2}
 -2 
 (\sum_{n} w_{n} x_{n})^{2} 
 + ( \sum_{n'} w_{n'} x_{n'} )^{2}  \  ]
\label{log_like_2a}
\end{equation}

\noindent 
\begin{equation}
=  
- {1 \over 2} \sum_{n'}( {1 \over \sigma_{n'}^{2} } )  [ \ 
 \sum_{n} w_{n} x_{n}^{2}
 -
 (\sum_{n} w_{n} x_{n})^{2} ]
\label{log_like_3}
\end{equation}
\noindent
yielding
\begin{equation}\fbox{$ \ \ \ 
\mbox{log} L^{(k)}_{\mbox{\small max}} = 
- {1 \over 2}  
[ \sum_{n'}( {1 \over \sigma_{n'}^{2} } ) ] \ 
\sigma^{2}_{X} \ \ \ 
 $}
\label{final_ml}
\end{equation}
\noindent 
where 
\begin{equation}
\sigma^{2}_{X} \equiv
 \sum_{n} w_{n} x_{n}^{2}
 -
 (\sum_{n} w_{n} x_{n})^{2} 
\end{equation}
is the \emph{weighted average 
variance} of the measured signal values in the block.
It makes sense that the block fitness function is
proportional to the negative of the variance: 
the best constant model for the block should have
minimum variance.

\subsection{Point measurements: 
Marginal Posterior, Flat Prior}
\label{point_2}

First, consider the simplest choice,
the flat, unnormalizable prior
\begin{equation}\label{flat_prior}
 P(\lambda) = P^{*}  \ \ \  (\mbox{for all values of} \  \lambda) \ ,
\end{equation}
\noindent
giving equal weight to all values. 
The marginal posterior for block $k$ is then, 
from Eq. (\ref{block_like}),
\begin{equation}  
 P^{k} = P^{*} 
  {(2 \pi ) ^{- { N_{k} \over 2} }  \over   \prod_{n} \sigma_{n}  } \ \
  { { \int_{-\infty}^{\infty} 
e^{ - {1 \over 2} \sum_{n} ({  x_{n} - \lambda \over \sigma_{n}} )^2 }
} }
d  \lambda
\end{equation}
\noindent
Using the definitions introduced
above in eqs. (\ref{aa}), (\ref{bb}), and (\ref{cc})
we have
\begin{equation}
P^{k}  = P^{*}
{(2 \pi ) ^{- { N_{k} \over 2} }  \over   \prod_{n} \sigma_{n}  } 
 \int_{-\infty}^{\infty}
e^{ - ( a_{k} \lambda^2 + b_{k} \lambda + c_{k} ) } \ d\lambda \ .
\label{flat_abc}
\end{equation}
\noindent
Using standard ``completing the square,'' letting 
$z = \sqrt{ a_{k} } ( \lambda + { b_{k} \over 2 a_{k} } )$, 
giving 
\begin{equation}
z^{2} = a_{k} (  \lambda + { b_{k} \over 2 a_{k} } ) ^{2} 
= a_{k} (  \lambda ^{2} + { \lambda  b_{k}  \over a_{k} } 
+  {b_{k}^{2}  \over 4 a_{k}^{2} } ) 
= a_{k}  \lambda ^{2} + b_{k} \lambda  + c_{k}
+  {b_{k}^{2}  \over 4 a_{k} }  - c_{k} \ ,
\end{equation}
\noindent
and then using 
\begin{equation}
 \int_{-\infty}^{+\infty}
e^{ - z^{2} } { dz \over \sqrt{a_{k} } } = \sqrt{ \pi \over a_{k} } \ .
\label{norm_integral}
\end{equation}
\noindent
we have
\begin{equation}
P^{k}  = P^{*}
{(2 \pi ) ^{- { N_{k} \over 2} }  \over   \prod_{n} \sigma_{n}  } 
\sqrt{ {\pi \over a_{k}}} e^{ ({b_{k}^{2} \over 4 a_{k} })  - c_{k}
}
\end{equation}
From this result, the log-posterior fitness function is 
\begin{equation}\fbox{$ \ \ \
log P^{k}_{\mbox{\small 0}} 
 - A_{k} =   \mbox{log}(  P^{*}\sqrt{ {\pi \over a_{k}  } })
+  ({b_{k}^{2} \over 4 a_{k} })  - c_{k} 
\ \ \ $}
\label{final_flat}
\end{equation}
\noindent 
where
\begin{equation}
A_{k} =  - {N_{k} \over 2} \mbox{log}( 2 \pi )
- \sum \mbox{log}( \sigma_{n} )
\label{a_def}
\end{equation}
and the subscript $0$ refers to the fact that
the marginal posterior was obtained with
the unnormalized prior.
The second and third terms in 
Eq. (\ref{final_flat} )
are invariant
under the transformation (\ref{trans}).
Further, since  the
integral of $P(\lambda)$ with respect to $\lambda$
must be dimensionless, we have 
$P^{*} \sim {1 \over \lambda}  \sim {1 \over x}$,
so $P^{*}$ and $\sqrt{a_{k}}$ have the
same $a$-dependence,
yielding a formal invariance for (\ref{final_flat}).
However the prior in eq. (\ref{flat_prior}) is not
normalizable, so that technically $P^{*}$
is undefined.
A way to make practical use of this
formal invariance is simply to include 
a constant $P^{*}$ that has the proper dimension
(${x^{-1}}$).

\subsection{Point Measurements:
Marginal Posterior,
Normalized Flat Prior}
\label{point_3}

Marginalizing the likelihood in 
eq. (\ref{block_like}) with the prior in
eq. (\ref{flat_prior_normalizable}), 
yields for the marginal posterior 
for block $k$:
\begin{equation}  
 P^{k} = P^{(\Delta)} 
  {(2 \pi ) ^{- { N_{k} \over 2} }  \over   \prod_{n} \sigma_{n}  } \ \
  { { \int_{\lambda_{1}}^{\lambda{2}} 
e^{ - {1 \over 2} \sum_{n} ({  x_{n} - \lambda \over \sigma_{n}} )^2 }
} }
d  \lambda
\end{equation}
As before
\begin{equation}
P^{k}  = P^{(\Delta)}
{(2 \pi ) ^{- { N_{k} \over 2} }  \over   \prod_{n} \sigma_{n}  } 
 \int_{\lambda_{1}}^{\lambda{2}}
e^{ - ( a_{k} \lambda^2 + b_{k} \lambda + c_{k} ) } \ d\lambda 
\end{equation}
\noindent
Now complete the square by letting 
$z = \sqrt{ a_{k} } ( \lambda + { b_{k} \over 2 a_{k} } )$, 
giving 
\begin{equation}
z^{2} = a_{k} (  \lambda + { b_{k} \over 2 a_{k} } ) ^{2} 
= a_{k} (  \lambda ^{2} + { \lambda  b_{k}  \over a_{k} } 
+  {b_{k}^{2}  \over 4 a_{k}^{2} } ) 
= a_{k}  \lambda ^{2} + b_{k} \lambda  
+  {b_{k}^{2}  \over 4 a_{k} } + c_{k} - c_{k}
\end{equation}
\noindent
so we have 
\begin{equation}
 P^{k}  = P^{(\Delta)}
{(2 \pi ) ^{- { N_{k} \over 2} }  \over   \prod_{n} \sigma_{n}  } 
e^{ ( {b_{k}^{2} \over 4 a_{k} }   - c_{k} )}
\int_{z_{1}}^{z_{2}}
e^{ - z^{2} } { dz \over \sqrt{a_{k} } } 
\label{fitness_erf}
\end{equation}
\noindent
where
\begin{equation}
z_{1,2}  = \sqrt{ a_{k} } ( \lambda_{1.2} + { b_{k} \over 2 a_{k} } )
\end{equation}
Finally, introducing the error function
\begin{equation}
\mbox{erf}( x ) = {2 \over \sqrt{ \pi } } \int_{0}^{x} e^{ - t^{2} } dt
\end{equation}
\noindent
we have
\begin{equation}
 P^{k}  = P^{(\Delta)} { \sqrt{ \pi } \over 2}
{(2 \pi ) ^{- { N_{k} \over 2} }  \over  \sqrt{a_{k} }  \prod_{n} \sigma_{n}  } 
e^{ ( {b_{k}^{2} \over 4 a_{k} }   - c_{k} )}
[\mbox{erf}( z_{2} ) - \mbox{erf}( z_{1} ) ]
\label{final_erf_1}
\end{equation}
\noindent
Taking the log gives the final expression
\begin{equation}\fbox{$ \ \ \ 
 {\mbox log}  P_{\Delta}^{k} - A_{k} 
 = 
 {\mbox log} ( P^{(\Delta)} \sqrt{ \pi \over a_{k}  } )
 + ( { b_{k}^{2} \over 4 a_{k} }   - c_{k} ) 
 +
{\mbox log} [{ \mbox{erf}( z_{2} ) - \mbox{erf}( z_{1} )  \over 2 }]
\ \ \ $}
\label{final_erf_2}
\end{equation}

\noindent
where the subscript $\Delta$ indicates the fact 
that this result is based on the finite-range
prior in eq. (\ref{flat_prior_normalizable}).
Note that this fitness function is
manifestly invariant under
the transformation in eq. (\ref{trans}),
for the same reasons discussed
at the end of the previous section, plus
the invariance of $z_{1,2}$.
In the limits 
$z_{1}  \rightarrow -\infty $
and
$z_{2}  \rightarrow \infty $,
$\mbox{erf}( z_{2} ) - \mbox{erf}( z_{1} ) \rightarrow 2$,
and we recover eq.(\ref{final_flat}) --
but remember that in this limit the invariance
is only formal.

\subsection{Point Measurements:
Marginal Posterior,  
Gaussian Prior}
\label{point_4}

Finally,
consider using the following 
normalized Gaussian prior for $\lambda$:
\begin{equation}\label{gauss_prior}
P(\lambda) =  {1 \over \sigma_{0} \sqrt{ 2 \pi } }
e^{- {1 \over 2} ( {\lambda - \lambda_{0} \over \sigma_{0}} ) ^{2} }
\label{normal_data_structure}
\end{equation}
\noindent
corresponding to prior knowledge 
that roughly speaking $\lambda$ most likely
lies 
in the range $~\lambda_{0} \pm \sigma_{0}$,
with a normal distribution.
This prior is not to be confused with
the Gaussian form for the likelihood
in eq. (\ref{likelihood}).

Eq.\ (\ref{block_like}), when $\lambda$ 
is marginalized with this prior,
becomes 
\begin{equation}
L^{(k)} = 
{1 \over \sigma_{0} \sqrt{ 2 \pi } }
[ {( 2 \pi )^{-({N_{k} \over 2})}  \over \prod_{n'} \sigma_{n'} } ] 
\int
e^{- {1 \over 2} 
[ \lambda^{2} ( {1 \over  \sigma_{0}^{2} }  +  \sum_{n}  {1 \over \sigma_{n}^{2}  }) 
+
\lambda ( - { 2 \lambda_{0}  \over \sigma_{0}^2 }  - { 2 x_{n} \over \sigma_{0}^2 } )
+ ( { \lambda_{0}^{2} \over \sigma_{0}^{2} } + \sum_{n} { x_{n}^{2} \over \sigma_{n}^{2} }) }
\label{block_like_3}
\end{equation}

\noindent 
so with
\begin{equation}
a_{k} = {1 \over 2} ( 
{1 \over  \sigma_{0}^{2}} + \sum_{n} {1 \over \sigma_{n}^{2}} )
\label{aaa}
\end{equation}

\noindent
\begin{equation}
b_{k} = - ( { \lambda_{0} \over \sigma_{0}^{2}}
+ \sum_{n} {x_{n} \over \sigma_{n}^{2}} )
\label{bbb}
\end{equation}

\noindent
and
\begin{equation}
c_{k} = {1 \over 2}  ( {\lambda_{0}^{2}  \over  \sigma_{0}^{2} } +
 \sum_{n} {x_{n}^{2} \over \sigma_{n}^{2}} )
 \label{ccc}
\end{equation}
\noindent
and eq. (\ref{flat_abc}) is recovered,
so that eq.(\ref{final_flat}), with the
redefined coefficients
in eqs. (\ref{aaa}), (\ref{bbb}) and (\ref{ccc}),
gives the final fitness function.

Any of the log fitness functions
in eqs. (\ref{final_ml}), (\ref{final_flat}),
or (\ref{final_erf_2})
can be used 
for the point measurement data mode 
in this section.
No general guidance for this
 depending on convenience
or the 
kind of prior information for the 
signal parameters that makes sense.

\subsection{Data with Dispersed Measurements}
\label{dispersed_data}

Throughout it has been presumed that 
two things are small compared to any relevant time scales:
errors in the determination of times of events,
and
the intervals over which individual 
measurements are obtained as averages.
These assumptions justify treatment
of the corresponding data modes as points
in  
 \S\ref{event_data}
 and
\S\ref{point_measurements}
respectively.
Below are discussions of 
data that are dispersed because of
(1) random errors in event times 
and 
(2) measurements that are summations or averages 
over non-negligible intervals.
Binned data, an example of the latter,
have already been treated in \S\ref{binned_event_data}
and are not discussed here.

A simple \emph{ad hoc} 
way to deal with both of these situations 
is to compute kernel functions for each data point,
representing the window or error distribution
in either of the two above contexts.
Each such function would be
centered at the corresponding measured value, 
evaluated at all of the data points,
and normalized to represent unit intensity.
Each such kernel would be maximum at 
the data point at which it is centered, but
distribute some weight to the other data cells.
The sum of all of these kernels 
would then be a set of weights 
at each measurement,
which could then be treated as ordinary
event data but with fractional 
rather than unit weights.
The \emph{ad hoc} aspect of this
approach lies in the way the fitness function
is extended.
The following sub-sections 
provide more rigorous analysis.

\subsubsection{Uncertain Event Locations}

Timing of events is always uncertain at some level.
Here we treat the case where the 
error distribution is wide enough
to make the point approximation inappropriate.
Rare for photon time series, 
with microsecond timing errors,
this situation is more common in other contexts 
and with other independent variables.
With overlapping error distributions 
even the order of events can be uncertain.
In the context described in \S\ref{histograms}
one often wants to construct histograms
from measurements with errors -- errors 
that may be different for each point (then called
\emph{heteroscedastic errors}).

A simple modification of the fitness function
described in \S\ref{event_data} addresses 
this kind of data.
On the right-hand side of 
Eq. (\ref{fitness_event})
$N^{(k)}$ quantifies 
the contribution of the individual events
within block $k$.
In extending the reasoning leading 
to this fitness function, 
the main issue concerns 
events with error distributions
that have fractional overlap 
with the extent of block $k$ -- for
events distributed entirely outside (inside) 
obviously contribute in no way (fully) to 
block fitness.
By the law for the sum of 
probabilities of independent events, 
in the log-likelihood implicit in Eqs. (\ref{cash})
and (\ref{cash_constant})
$N^{(k)}$
is replaced by the sum of the 
areas under the probability distributions 
overlapping block $k$, namely 
$\sum _{i \in k} p^{(i)}$ 
summed over all events with significant contribution to block $k$,
and $p^{(i)}$ is the integral of the overlapping part of
the error distribution, 
a fraction between $0$ and $1$.
Thus we have 
\begin{equation}
\label{distributed_events}
{\mbox log} L^{(k)}(\lambda)  =  {\mbox log} \lambda 
\sum _{i \in k} p^{(i)}
 \ - \lambda T^{(k)} \ 
\end{equation}
\noindent
in place of Eq. (\ref{fitness_event}), 
with the analogous constant term 
on the left-hand side of that equation dropped.
This result holds because 
a given datum falling 
inside and outside a block 
are mutually exclusive events.

Implementing this relationship 
in the algorithm is easily accomplished.
For a given event and the interval assigned to it
(\emph{cf.} Figure \ref{voronoi_cell}
in \S\ref{poisson_nature}) 
sum the overlap fractions with that interval 
of all events
-- including that event itself.
These quantities could be approximated 
with very simple or complex quadrature 
schemes, depending on the context and
the way in which the relevant distributions
are represented.
Normally the array \verb+nn_vec+, as in the
code fragment in \S\ref{appendix_a}, is all 1's
(or counts of events with identical time-tags there are any);
but here replace it with these summed event weights.
This construction automatically assigns the
correct fractional weights to the block with
no further alteration of the algorithm.

\subsubsection{Measurements in Extended Windows}

This section discusses the case of 
\emph{distributed measurements} 
in the sense that the time of measurement
is either uncertain or is effectively an interval 
rather than a point.
(This is different from the 
use of this term 
in \S\ref{point_measurements}
to describe the
distribution of the measurement error
in the dependent variable.)
Measurements may refer to a quantity
averaged over a range of values of $t$,
not at a single time as in 
\S\S
\ref{point_measurements}, 
\ref{point_1}, 
\ref{point_2}, \ref{point_3}, and \ref{point_4}.
In the context of histograms (\S\ref{histograms})
the measured quantity becomes the independent variable,
and the dependent variable is
an indicator marking the presence of 
the measurement there.
In both cases the measurement 
can be thought of as distributed over
an interval, not just at a point.

In this case the data cell array would be 
augmented by the inclusion of a window
function, 
indicating the variation of the instrumental sensitivity:
\begin{equation}
x = \{ x_{n}, t_{n}, w_{n}(t - t_{n}) \} \ \ n = 1, 2, \dots , N \ ,
\end{equation}
\noindent
where $w_{n}(t)$
describes, for the value reported as $X_{n}$,
the relative weights assigned to
times near $t_{n}$.

This is a nontrivial complication if the
window functions overlap,
but can nevertheless be handled with the
same technique.

We assume the standard piece-wise constant model of the underlying signal,
that is, a set of contiguous blocks:
\begin{equation}
B(x) = \sum_{j=1}^{N_{b}} B^{(j)}( x )
\end{equation}
where each block is represented as a {\it boxcar} function:
\begin{eqnarray}
 B^{(k)}( x ) = \{
             \begin{array}{ll}
             B_{j} & \zeta_{j} \le x  \le  \zeta_{j+1}\cr
             0     & \mbox{otherwise}
             \end{array}
\end{eqnarray}
\noindent
the $\zeta_{j}$ are the change-points, satisfying
\begin{equation}
min( x_{n} ) \le \zeta_{1} \le \zeta_{2} \le \dots
   \zeta_{j} \le \zeta_{j+1} \le \dots \le \zeta_{N_{b}} \le max( x_{n} )
\end{equation}
\noindent
and the $B_{j}$ are the heights of the blocks.

The value of the observed quantity, $y_{n}$, at $x_{n}$,
under this model is
\begin{eqnarray}
\begin{array}{ll}
\hat{y}_{n}  & = \int w_{n}(x) B( x ) dx \cr
 \           & = \int w_{n}(x) \sum_{j=1}^{N_{b}} B^{(j)}( x ) dx \cr
 \           & = \sum_{j=1}^{N_{b}} \int w_{n}(x) B^{(j)}( x ) dx \cr
 \           & = \sum_{j=1}^{N_{b}} B_{j} \int_{\zeta_{j}}^{\zeta_{j+1}} w_{n}(x) dx
\end{array}
\end{eqnarray}
\noindent
so we can write
\begin{equation}
\hat{y}_{n} = \sum_{j=1}^{N_{b}} B_{j} G_{j}(n)
\end{equation}
\noindent
where
\begin{equation}
G_{j}(n) \equiv \int_{\zeta_{j}}^{\zeta_{j+1}} w_{n}(x) dx
\end{equation}
\noindent is the inner product of the $n$-th weight function with
the support of the $j$-th block.  The analysis in \cite{brett_1}
shows how do deal with the non-orthogonality that is generally the
case here.\footnote{If the weighting functions are delta functions,
it is easy to see that $G_{j}(n)$ is non-zero if and only if $x_{n}$
lies in block $j$, and since the blocks do not overlap the product
$G_{j}(n) G_{k}(n)$ is zero for $j \ne k$, yielding  orthogonality,
$\sum _{N} G_{j}(n) G_{k}(n) = \delta_{j,k}$. And of course there
can be some orthogonal blocks, for which there happens to be no
``spill over'', but these are exceptions.}

The averaging process in this data model induces dependence among
the blocks. The likelihood, written as a product of likelihoods of
the assumed independent data samples, is
\begin{eqnarray}
P( \mbox{Data} | \mbox{Model} ) & = \prod_{n=1}^{N} P( y_{n} | \mbox{Model} ) \\
                                & = \prod_{n=1}^{N} {1 \over \sqrt{ 2 \pi \sigma_{n}^{2} }}
e^{- {1 \over 2} ({ y_{n} - \hat{y}_{n} \over \sigma_{n} } ) ^{2} } \\
                                & = \prod_{n=1}^{N} {1 \over \sqrt{ 2 \pi \sigma_{n}^{2} }}
e^{- {1 \over 2} ({ y_{n} - \sum_{j=1}^{N_{b}} B_{j} G_{j}(n)  \over \sigma_{n} } ) ^{2} } \\
           & = Q
e^{- {1 \over 2} ({ y_{n} - \sum_{j=1}^{N_{b}} B_{j} G_{j}(n)
\over \sigma_{n} } ) ^{2} } \ ,
\end{eqnarray}
\noindent
where
\begin{equation}
Q \equiv \prod_{n=1}^{N} {1 \over \sqrt{ 2 \pi \sigma_{n}^{2} }} \ .
\end{equation}
After more algebra
and adopting a new notation, symbolized by
\begin{equation}
{ y_{n} \over \sigma_{n}^{2} } \rightarrow y_{n}
\end{equation}
and
\begin{equation}
{ G_{k}(n) \over \sigma_{n}^{2} } \rightarrow G_{k}(n) \ ,
\end{equation}
\noindent
we arrive at
\begin{equation}
log P( \{y_{n}\} | B ) = Q e^{- {H \over 2}} \ ,
\end{equation}
\noindent
where
\begin{equation}
H \equiv
\sum_{n=1}^{N} y_{n}^{2}
- 2 \sum_{j=1}^{N_{b}}  B_{j} \sum_{n=1}^{N} y_{n} G_{j}(n)
+ \sum_{j=1}^{N_{b}}  \sum_{k=1}^{N_{b}}
 B_{j}  B_{k}
\sum_{n=1}^{N}
  G_{j}(n)  G_{k}(n) \ .
\end{equation}
\noindent
The last two equations
are equivalent to Eqs. (3.2) and (3.3) of \cite{brett_1},
so that the orthogonalization of the basis functions
and the final expressions follow exactly as in that reference.

\subsection{Piecewise Linear Model: Event Data}
\label{model_linear_fitness}

Here we outline the computations 
of a fitness function for the piecewise linear model
in the case of event data.
This means that the event rate for a block 
is assumed to be linear, as in Eq. (\ref{model_linear}).

For convenience we take the fiducial time $t_{\mbox{fid}}$
to be $t_{2}$, the time
at the end of the block.
Take $t_{1}$ to be the time at the beginning,
so $M = t_{2} - t_{1}$ is the length of the block, and
the signal $x$ is $\lambda  ( 1 - aM )$ at the beginning of
the block and  $\lambda $ at the end,
and varies linearly in between.

The block likelihood for the case of event data $t_{i}$
is
\begin{equation}
L( \lambda, a ) = \sum_{i=1}^{N_{k}} {\mbox log} 
[ \  \lambda  ( 1 + a ( t_{i} - t_{2} )  ) \  ]
- \int_{t_1} ^{t_2} \lambda  ( 1 + a ( t - t_{2} )  )  dt
\end{equation}
\noindent
where the sum is over the $N_{k}$ events in the block
and
the integral is over the time interval covered by the block.
Simplifying we have
\begin{equation}
L( \lambda, a ) = N_{k} \ {\mbox log} \lambda 
+ \sum_{i=1}^{N_{k}} {\mbox log} 
[ \  ( 1 + a ( t_{i} - t_{2} )  ) \  ]
-  \lambda 
[(1 - a t_{2} ) t  + {a \over 2} t^{2} ]_{t_1}^{t_2}
\end{equation}
\begin{equation}
L( \lambda, a ) = N_{k} \ {\mbox log} \lambda 
+ \sum_{i=1}^{N_{k}} {\mbox log} 
[ \  ( 1 + a ( t_{i} - t_{2} )  ) \  ]
-  \lambda M_{k} ( 1- {a \over 2} M_{k} )
\end{equation}

Now let's compute the maximum likelihood as a 
function of $\lambda$ and $a$, starting
by setting 
\begin{equation}
{\partial L \over \partial \lambda } = 
{ N_{k} \over \lambda } 
-  M_{k} ( 1- {a \over 2} M_{k} ) = 0
\end{equation}
\noindent
so that at the maximum of this likelihood we have
\begin{equation} 
\lambda =
{ N_{k} \over M_{k} ( 1- {a \over 2} M_{k} ) }
\end{equation}
\noindent
and therefore
\begin{equation}
L( \lambda_{max}, a ) = N_{k} \ {\mbox log}[ { N_{k} \over M_{k} ( 1- {a \over 2} M_{k} ) } ]
+ \sum_{i=1}^{N_{k}} {\mbox log} 
[ \  ( 1 + a ( t_{i} - t_{2} )  ) \  ]
- N_{k} 
\end{equation}
\noindent
\begin{equation}
{\partial L \over \partial a } = 
N_{k} \ {\mbox log}[ { N_{k} \over M_{k} ( 1- {a \over 2} M_{k} ) } ]
+ \sum_{i=1}^{N_{k}} {\mbox log} 
[ \  ( 1 + a ( t_{i} - t_{2} )  ) \  ]
- N_{k} 
\end{equation}
\noindent
\begin{equation}
{\partial L \over \partial a } = 
\sum_{i=1}^{N_{k}} { ( t_{i} - t_{2} )  
 \over  1 + a ( t_{i} - t_{2} )   }
+ { \lambda \over 2} M_{k}^{2} = 0
\end{equation}
\noindent
\begin{equation}
{1 \over N_{k} }
\sum_{i=1}^{N_{k}} { ( t_{i} - t_{2} )  
 \over  1 + a ( t_{i} - t_{2} )   }
+  { {1 \over 2} M_{k}  \over ( 1- {a \over 2} M_{k} ) }   = 0
\end{equation}
\noindent
\begin{equation}
f( a ) = {1 \over N_{k} }
\sum_{i=1}^{N_{k}} { ( t_{i} - t_{2} )  
 \over  1 + a ( t_{i} - t_{2} )   }
+  { {1 \over 2} M_{k}  \over ( 1- {a \over 2} M_{k} ) }  
\end{equation}
\noindent
\begin{equation}
f'( a ) = -{1 \over N_{k} }
\sum_{i=1}^{N_{k}} { ( t_{i} - t_{2} ) ^{2} 
 \over  [1 + a ( t_{i} - t_{2} ) ] ^{2} }
-  { {1 \over 4} M_{k}^{2}  \over ( 1- {a \over 2} M_{k} )^{2} }  
\end{equation}

\begin{equation}
\lambda = - {2 \over M_{k}^{2} }
\sum_{i=1}^{N_{k}} { ( t_{i} - t_{2} )  
 \over  1 + a ( t_{i} - t_{2} )   }
\end{equation}
\noindent
\begin{equation}
{ N_{k} \over ( 1- {a \over 2} M_{k} ) }
 = - {2 \over M_{k} }
\sum_{i=1}^{N_{k}} { ( t_{i} - t_{2} )  
 \over  1 + a ( t_{i} - t_{2} )   }
\end{equation}
\noindent
\begin{equation}
{  ( 1- {a \over 2} M_{k} ) \over   N_{k}  }
 = - {M_{k}   \over 2 \sum_{i=1}^{N_{k}} { ( t_{i} - t_{2} )  
 \over  1 + a ( t_{i} - t_{2} )   } }
\end{equation}
\noindent
\begin{equation}
1- {a \over 2} M_{k} 
 = - {1 \over 2} M_{k}  N_{k}   ( \sum_{i=1}^{N_{k}} { ( t_{i} - t_{2} )  
 \over  1 + a ( t_{i} - t_{2} )   } )^{-1}
\end{equation}
\noindent
\begin{equation}
a  = {2 \over  M_{k} }  +  
N_{k}  
( \sum_{i=1}^{N_{k} }
{ ( t_{i} - t_{2} )  \over  1 + a ( t_{i} - t_{2} )  }  )^{-1}
\end{equation}
\noindent

\subsection{Piecewise Exponential Model: 
Event Data}
\label{model_exponential_fitness}

In this case we model the signal as varying
exponentially across the time interval contained in the block.
Denoting the times 
beginning and ending the block
as $t_{1}$ and $t_{2}$,
and taking 
the latter as the fiducial time in 
Equation (\ref{model_exponential}),
the signal is $\lambda  e^{ -aM }$ at the beginning of
the block and  $\lambda $ at the end.

Much as in \S\ref{model_linear_fitness}
the block likelihood for the case of event data $t_{i}$
is the follow expression involving 
a sum over the $N_{k}$ events in the block
and an
integral over the time interval covered by the block:
\begin{equation}
L( \lambda, a  ) = \sum_{i=1}^{N_{k}} {\mbox log} [ \  \lambda  e^{ a ( t_{i} - t_{2} ) }\  ]
- \int_{t_1} ^{t_2} \lambda  e^{ a ( t - t_{2} ) } dt
\end{equation}
\noindent
\begin{equation}
L( \lambda, a  ) = N_{k} \  {\mbox log} \lambda + 
a \sum_{i} { ( t_{i} - t_{2} ) } \  
- \lambda  ( 
{ 1 - e^{ -a M }  \over a } ) 
\end{equation}
\noindent
where $M = t_{2} - t_{1}$ is the length of the block.

Now let's compute the maximum likelihood as a 
function of $\lambda$ and $a$:
\begin{equation}
{\partial L \over \partial \lambda } = 
{ N_{k} \over \lambda } 
- ( 
{ 1 - e^{ -a M }  \over a } ) 
\end{equation}
\noindent
and therefore at the maximum we have
\begin{equation}
 \lambda  = 
{ a  N_{k} \over  1 - e^{ -a M }  }
\end{equation}

\begin{equation}
{\partial L \over \partial a } = 
\sum_{i} { ( t_{i} - t_{2} ) } \  
- [  N_{k} ( 1 - e^{ -a M })^{-1} ]   [  ( M  +  a^{-1}) e^{ -a M } - a^{-1} ]
\end{equation}
\noindent
\begin{equation}
L_{max} ( a ) = N_{k} \  {\mbox log} ({ a  N_{k} \over  1 - e^{ -a M }  } ) + 
a \sum_{i} { ( t_{i} - t_{2} ) } \  
- { a  N_{k} \over  1 - e^{ -a M }  } ( 
{ 1 - e^{ -a M }  \over a } ) 
\end{equation}
\noindent
\begin{equation}
L_{max} ( a ) = N_{k} \  {\mbox log} ({ a  N_{k} \over  1 - e^{ -a M }  } )+ 
a \sum_{i} { ( t_{i} - t_{2} ) } \  
- N_{k} 
\end{equation}
\noindent

\begin{equation}
{\partial L_{max}( a ) \over \partial a } = 
 N_{k} \  
({ 1 - e^{ -a M }   \over  a  N_{k} } )
Q + 
\sum_{i} { ( t_{i} - t_{2} ) }  
\end{equation}
\noindent
where 
\begin{equation}
Q 
=  N_{k} [ ( 1 - e^{ -a M}  )^{-1} - a ( 1 - e^{ -a M}  )^{-2} M e^{-aM}  ] 
\end{equation}

\noindent
\begin{equation}
{\partial L_{max}( a ) \over \partial a } = 
{N_{k} \over a } -
M N_{k} { e^{-aM} \over ( 1 - e^{ -a M}  )}
+ 
\sum_{i} { ( t_{i} - t_{2} ) }  
\end{equation}
To solve for the value of $a$ that makes this
derivative zero (to find the maximum of the likelihood)
we will use Newton's method to find the zeros of  
\noindent
\begin{equation}
f(a) = {\partial L_{max}( a ) \over \partial a } / N_{k}  =
{1 \over a } -
M e^{-aM} ( 1 - e^{ -a M}  )^{-1}
+ S
\end{equation}
\noindent 
where
\begin{equation}
S ={1 \over N_{k}}  \sum_{i} { ( t_{i} - t_{2} ) }  
\end{equation}
\noindent
is the mean of the differences
between the event times and the time at the end of the block.
The iterative equation is
\begin{equation}
a_{k+1} = a_{k} - { f(a_{k} ) \over f'(a_{k} ) }
\end{equation}
\noindent
and since $S$ is a constant we have 
\begin{equation}
f'(a) = 
- {1 \over a^{2} } -
M [
 -M e^{-aM} ( 1 - e^{ -a M}  )^{-1} 
 -
 M e^{-aM} ( 1 - e^{ -a M}  )^{-2} e^{ -a M}]
\end{equation}
\begin{equation}
f'(a) = 
- {1 \over a^{2} } +
M^{2} e^{-aM} ( 1 - e^{ -a M}  )^{-1}   [
 1
 +
 e^{-aM} ( 1 - e^{ -a M}  )^{-1}]
\end{equation}
\noindent
and defining
\begin{equation}
Q(a) = e^{-aM} ( 1 - e^{ -a M}  )^{-1}
\end{equation}
\noindent
we have
\begin{equation}
f'(a) = 
- {1 \over a^{2} } +
M^{2} Q(a)  [1 + Q(a)]
\end{equation}
\noindent
and
\begin{equation}
a_{k+1} = a_{k} - { 
a_{k}^{-1} - M Q(a_{k}) + S
 \over 
 - a_{k}^{-2} + M^{2} Q(a_{k})  [1 + Q(a_{k}) ] ) 
 }
\end{equation}
\noindent

\newpage

\end{document}